\PassOptionsToPackage{numbers, compress}{natbib}
\documentclass{article}

\usepackage[preprint]{neurips_2026}

\usepackage[utf8]{inputenc}
\usepackage[T1]{fontenc}
\usepackage{hyperref}
\usepackage{url}
\usepackage{booktabs}
\usepackage{amsfonts}
\usepackage{amsmath}
\usepackage{amssymb}
\usepackage{nicefrac}
\usepackage{microtype}
\usepackage[table]{xcolor}
\usepackage{graphicx}
\usepackage{wrapfig}
\usepackage{listings}
\usepackage[most]{tcolorbox}
\usepackage[font=footnotesize,labelfont=bf]{caption}
\usepackage{enumitem}
\usepackage{cleveref}

\definecolor{citecolor}{HTML}{0071bc}
\definecolor{cadmiumgreen}{rgb}{0.0, 0.42, 0.24}
\definecolor{cornellred}{rgb}{0.7, 0.11, 0.11}

\definecolor{view0col}{RGB}{245,235,224}   
\definecolor{view21col}{RGB}{214,204,194}  
\newcommand{\winzero}[1]{#1}
\newcommand{\winseven}[1]{\cellcolor{view0col}#1}
\newcommand{\wintwentyone}[1]{\cellcolor{view21col}#1}
\newcommand{\winbase}[1]{#1}                        
\newcommand{\winretry}[1]{\cellcolor{view21col}#1}  

\hypersetup{
  colorlinks = true,
  linkcolor  = cornellred,
  citecolor  = citecolor,
  urlcolor   = cadmiumgreen,
}

\tcbset{
  reqbox/.style={
    breakable, enhanced,
    colback=teal!04, colframe=teal!25,
    coltitle=black, fonttitle=\bfseries\color{black},
    sharp corners, boxrule=0.6pt,
    left=1.6mm, right=1.6mm, top=1.2mm, bottom=1.2mm,
    before skip=6pt, after skip=6pt,
    fontupper=\footnotesize
  }
}

\definecolor{bpkey}{RGB}{45,45,45}
\definecolor{bpval}{RGB}{125,40,40}
\definecolor{bpcom}{RGB}{60,118,61}
\definecolor{bpbg}{RGB}{248,249,250}
\definecolor{bprule}{RGB}{200,200,200}

\lstdefinelanguage{blueprint}{%
  morekeywords={parts,construction_units,constructive_primitives,acceptance_claims,%
    role,envelope,must_fit_inside,bounding_envelope,support_zones,plane,footprint,%
    op,axis,radius_outer,wall_thickness,sweep_path,edge_selector,radius,id,metric,%
    operator,value,name,polygon_2d,extrude_axis,extrude_length,center_2d,%
    assembly_schema_version,geometry_definition,metadata,brief_id,units,%
    coordinate_system,material,yield_strength_MPa,safety_factor,additive,modifier,%
    subtractive,normal,offset,x_span,z_span,bounds,x,y,z,cylinder,fillet_hint,%
    extrude_polygon,subtract_cylinder,subtract_box,box_bounds,annular_sector,%
    pattern_radial},%
  morestring=[b]",%
  morecomment=[l]\#,%
  sensitive=true%
}

\newcommand{\imagefeedback}{rich-view image judge}

\newcounter{briefex}
\renewcommand{\thebriefex}{\arabic{briefex}}

\newcommand{\newbench}{Hephaestus-CCX}

\title{Self-Improving CAD Generation Agents with Finite Element Analysis as Feedback}

\author{
Guijin Son$^{1,2}$\thanks{Equal contribution.}
\quad
Jehyun Park$^{3}$\footnotemark[1]\hspace{0.35em}\thanks{Work was done while Jehyun was an intern at Seoul National University.}
\quad
Seyeon Park$^{4}$
\quad
Sunghee Ahn$^{1}$
\quad
\textbf{Youngjae Yu}$^{1}$
\\ \\
Seoul National University$^{1}$
\qquad
OneLineAI$^{2}$ 
\qquad
Sungkyunkwan University$^{3}$
\qquad \\
Ewha Womans University$^{4}$
\\ \\
\texttt{guijin.son@snu.ac.kr}
\qquad
\texttt{jaheon555@g.skku.edu}
}

\begin{document}

\maketitle

\begin{abstract}
Computer-aided design (CAD) is the backbone of modern industrial design, yet learned CAD generators still fall short of real engineering pipelines: they neither iterate like engineers nor evaluate what engineering requires. Prior work has treated CAD generation as two disjoint steps, part synthesis and assembly, where the former is graded by proximity to a gold reference and the latter, when handled at all, is reduced to a separate constraint solving step. In this work, we introduce a more industry-native task formulation that requires a model to produce a fully assembled multi-part STEP file from a free-form engineering brief, which is then validated via finite element analysis (FEA). FEA validation reveals that Codex (GPT-5.5) and Claude Code (Opus-4.7) agents do not produce a single strict-passing artifact in the main first-attempt sweep, with the best configuration meeting only about 20\% of typed requirements on average. Moreover, we introduce two additional supervision signals, a novel text-only blueprint schema and a 21-view image renderer that aids the agent's visual inspection, that better align the generation loop with how engineers iterate in practice. On \texttt{S2O} and \texttt{Fusion\,360}, the same feedback tools improve geometric reconstruction, with \texttt{GPT-5.5/xhigh} rising from 0.444 to 0.592 Box-IoU on \texttt{S2O} and from 0.397 to 0.505 on \texttt{Fusion\,360}. Together these signals move CAD programs toward artifacts that are not only visually plausible but also checked against physical and structural requirements.
\end{abstract}

\begin{figure*}[h]
\vspace{-1.25em}
\centering
\includegraphics[width=0.87\textwidth]{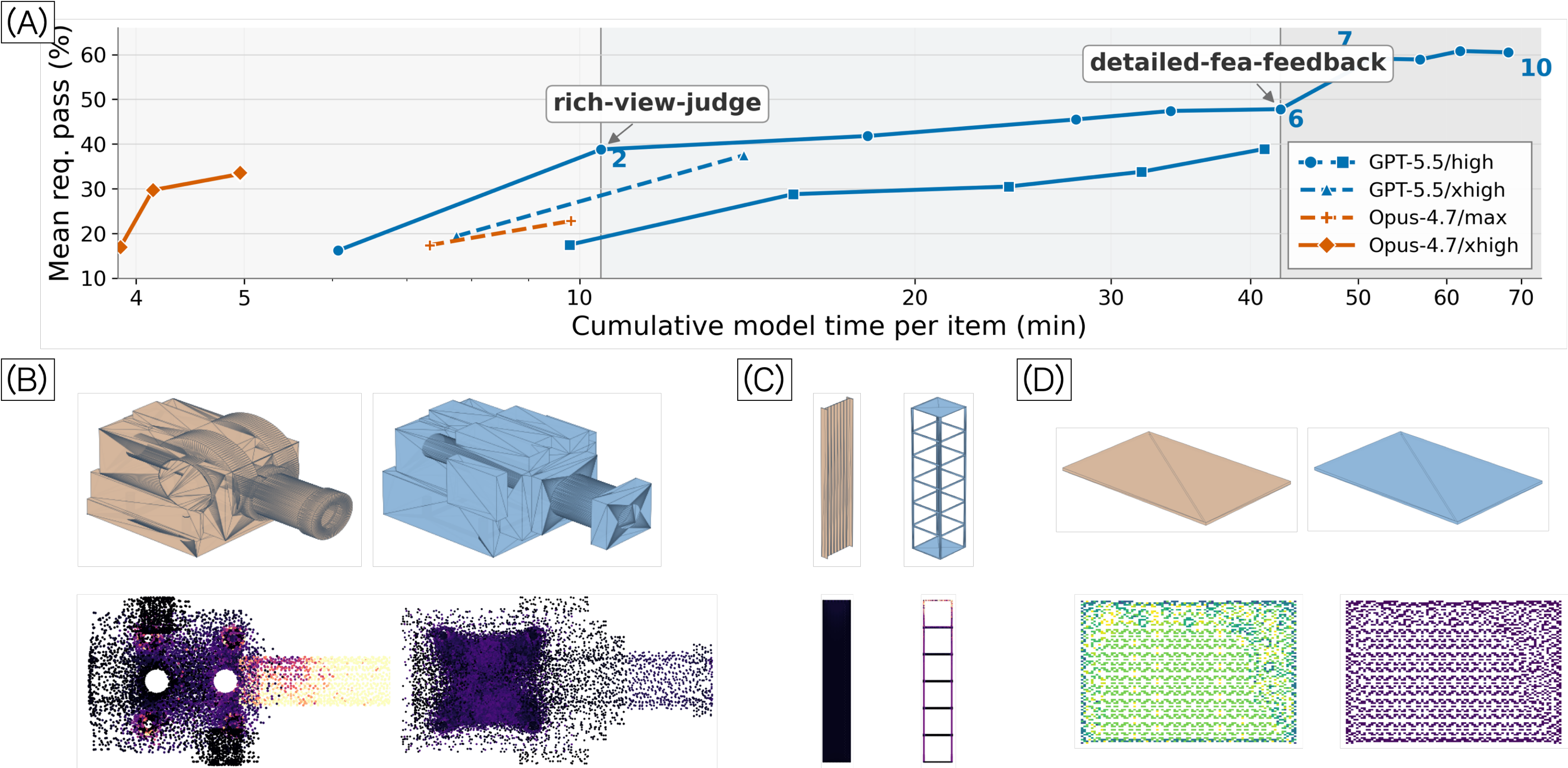}
\caption{\footnotesize \textbf{Repeated engineering feedback turns test-time compute into better CAD artifacts.} (A) Longer feedback loops steadily improve partial credit on \newbench{} as average model time scales from about 10 minutes to nearly 70 minutes per item. Mean requirement pass is the average per-case fraction of typed requirements satisfied. See~\Cref{sec:scaling} for scaling setup and qualitative details.
 (B--D) Successful retries show three repair modes. The launcher(B) removes fragile over-detailed geometry and routes load through a simpler stiff body. The column (C) adds bracing and section stiffness to recover axial capacity. The spacecraft panel (D) keeps a similar exterior shape, but fixes hidden density and mass-property errors.
}
\label{fig:compute-scaling}
\vspace{-1.75em}
\end{figure*}

\section{Introduction}
\label{sec:intro}

Recent learned CAD systems have made substantial progress in text-to-part generation, CAD-code synthesis, and assembly~\citep{text2cad,cadllama,ttcq,cadrille,cadassistant,joinable,automate}. These systems show that large models can translate natural language into executable modeling programs and plausible geometry. However, the dominant formulation remains weakly coupled to engineering validity. Outputs are commonly graded by distance to a reference shape~\citep{deepcad,cadsignet,cadcoder,cadrecode}, rendered visual plausibility~\citep{cadcodeverify,caddesigner,text2cad,evocad,seekcad}, topological validity~\citep{cadmium,cadsmith}, or mate prediction between parts that already exist~\citep{joinable,automate,fusion360gallery}. These signals miss failures that make a design unusable, such as a misplaced interface, insufficient clearance, an invalid load path, or a selector that cannot support downstream analysis. We therefore reframe CAD generation as an iterative tool-using process. An LLM agent writes a CadQuery program, executes it to export a STEP artifact, receives structured feedback from rendering, validation, and simulation tools, and revises the program and selector metadata before the next attempt. Our loop adds two agent-side tools, a structured blueprint and rich-view visual inspection, together with finite-element feedback from CalculiX~\citep{calculix}.

We equip this loop with a pre-CAD blueprint stage. The blueprint records the design commitments that the agent must satisfy when writing CAD. It constrains geometry to auditable parametric primitives, fixes envelopes and interfaces before code is emitted, and exposes dimensional and functional claims for validators and retry feedback. Once the agent generates CAD from this blueprint, the \imagefeedback{} renders the STEP from 21 calibrated views, including exterior views, close-ups, and internal x-ray cuts. This is a large increase over the small 4--6 render sets commonly used in visual CAD-code evaluations~\citep{cadcodeverify,caddesigner,text2cad,evocad,seekcad}, and it is meant to give the agent the static equivalent of walking around the assembly, zooming into interfaces, and taking section cuts. The agent can use these image reviews to fix visible geometry, assembly, and selector errors before final submission. Once the agent reports that the design is ready, an external FEA loop performs the analysis step that an engineer would run after inspection. It meshes the candidate design, runs CalculiX, and returns typed failures over stress, displacement, modal, buckling, contact, and clearance requirements. In our setting, the agent is prompted to consume these blueprint, visual, and FEA reports as engineering feedback, revise the CadQuery program and selector metadata, and resubmit a STEP artifact for up to 10 attempts. To evaluate this setting, we introduce \texttt{\newbench{}} (H-CCX), a benchmark of 50 engineering briefs collected from patents, supplier datasheets, engineering standards, regional industrial catalogs, and engineering competitions, each paired with executable requirement checkers. Each case asks for an assembled STEP artifact and is graded by whether the generated CAD satisfies the stated physical and geometric contract, not by whether it matches one reference mesh.

Our experiments show that this task remains nearly unsolved even for current frontier models. In the main Codex and Claude Code sweep, 400 first attempts do not produce a single strict-passing artifact, and one FEA-feedback round adds only one strict pass across another 400 revised submissions. Notably, partial-credit metrics show that the tools still move models in the right direction: 21-view feedback raises \texttt{GPT-5.5} from 19.4\% to 29.3\% mean requirement pass on \newbench{} and from 0.397 to 0.505 IoU on \texttt{Fusion\,360}, while blueprinting raises its \texttt{S2O} IoU from 0.444 to 0.592. Repeating the feedback loop compounds these gains. In our longest \texttt{GPT-5.5/high} run, the model spends 68 minutes per item on average, nearly a $7\times$ increase over the 10-minute two-attempt setting, and mean requirement pass rises from 38.8\% to 60.5\% with 9/50 strict-passing artifacts. This suggests that test-time compute can scale stably when it is organized as structured engineering feedback, not simply when it is spent as a larger one-shot reasoning budget.

This paper makes two contributions.

\begin{itemize}[leftmargin=2em]
\item \textbf{An engineering-grounded CAD-agent task.} We move CAD generation toward assembled STEP artifacts that are judged by geometric checks and finite-element analysis requirements, not only by reference matching or visual plausibility. To support future work on this setting, we release \newbench{}, a 50-case benchmark of single-part and multi-part engineering briefs paired with CalculiX evaluation kits and typed pass/fail requirement checkers.
\item \textbf{A study of feedback for CAD agents.} We implement structured blueprints, rich-view visual inspection, and FEA retry feedback inside production coding-agent harnesses. We measure where each feedback source improves frontier model performance, and how repeated feedback-driven repair compounds these gains over time.
\end{itemize}

\begin{figure*}[t]
\centering
\includegraphics[width=\textwidth]{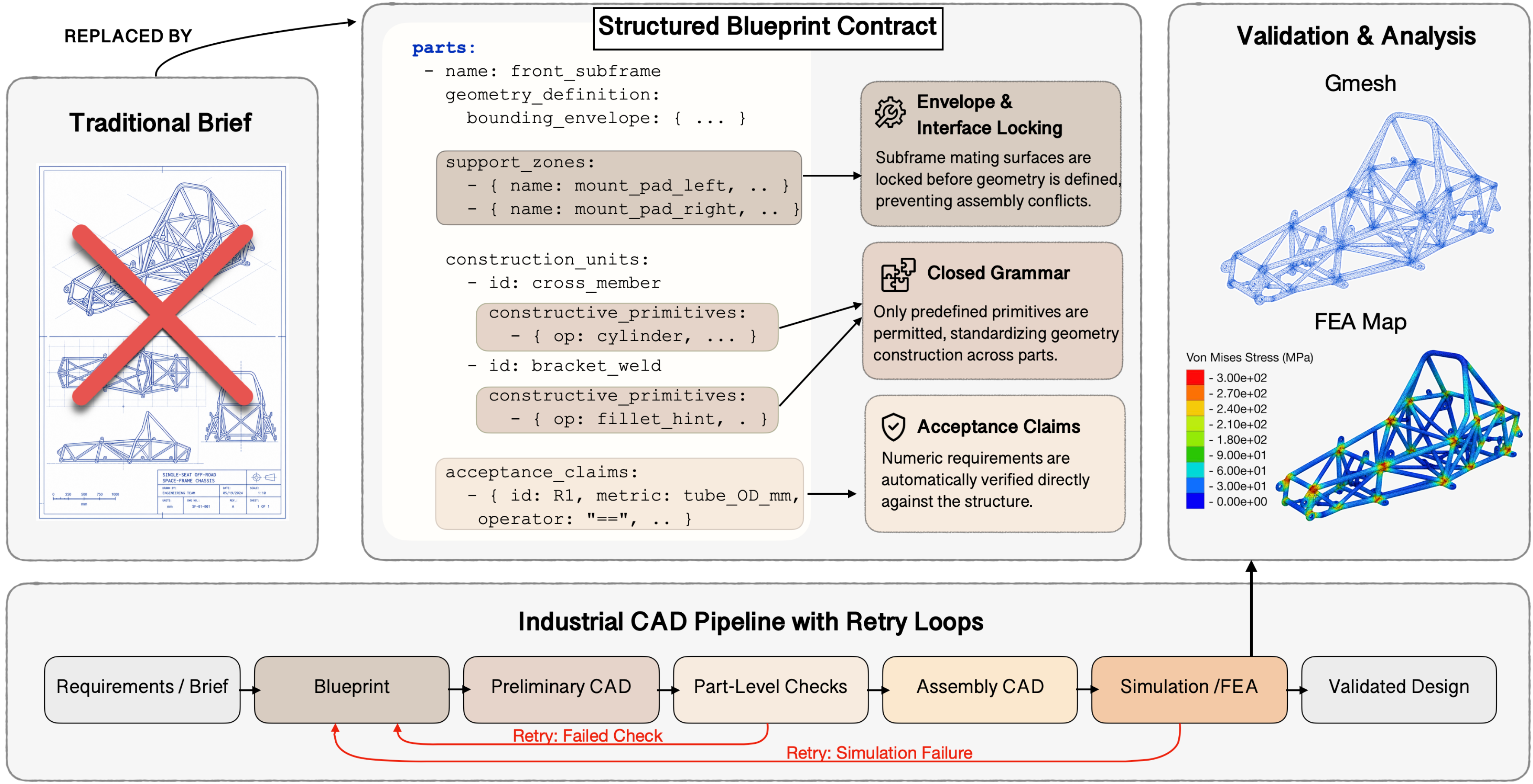}
\caption{\footnotesize \textbf{Overview of the CAD-agent pipeline.} A free-form engineering brief is converted into an optional schema-v4 blueprint, decomposed into construction units, assembled into a STEP artifact by a deterministic controller, and revised using rich-view inspection and FEA feedback. The controller owns execution, measurement, composition, and validation, while the agent owns design decisions and CAD-code repair.}
\label{fig:main}
\end{figure*}

\section{Background and Problem Formulation}
\label{sec:problem}

\subsection{Prior Work on CAD Generation}
\label{sec:problem:prior}

\paragraph{Gold reference matching.}
A long line of work casts CAD generation as a sequence-to-geometry problem and grades success by how close the output sits to a curated gold reference~\citep{deepcad,cadsignet,cadcoder,cadrecode}. Reported metrics are nearly identical across this body of work. Let $P$ and $Q$ denote point clouds sampled from the generated and reference solids, $A$ and $B$ their voxelized occupancy grids, and $N$ the number of generated programs. The four most widely used measures are Chamfer Distance (CD), F-score at threshold $\tau$, Intersection over Union (IoU), and Invalidity Ratio (IR), following the point-set metrics of \citet{fan2017point} and \citet{tatarchenko2019what}:
\begin{align*}
\mathrm{CD}(P,Q) &= \tfrac{1}{|P|}\sum_{p\in P}\min_{q\in Q}\|p-q\|_2^2 \;+\; \tfrac{1}{|Q|}\sum_{q\in Q}\min_{p\in P}\|p-q\|_2^2, \\
F_\tau &= \tfrac{2\, P_\tau R_\tau}{P_\tau + R_\tau}, \quad
\mathrm{IoU}(A,B) = \tfrac{|A \cap B|}{|A \cup B|}, \quad
\mathrm{IR} = \tfrac{|\{i : \text{program}_i \text{ does not yield a valid solid}\}|}{N},
\end{align*}
where $P_\tau$ and $R_\tau$ are the precision and recall of generated points lying within radius $\tau$ of the reference (and symmetrically). While works differ in how they represent the design (e.g., command tokens in \citet{cadllama}, CadQuery Python in \citet{ttcq,cadrille}, FreeCAD scripts in \citet{cadassistant}, and Blender scripts in \citet{threedgpt}), evaluation collapses to a single question: how close does the output land to the reference solid? Recent works do make progress on this front, including VLM judges on rendered images~\citep{cadcodeverify,caddesigner,text2cad,evocad,seekcad}. Other systems add topological validity checks~\citep{cadmium,cadsmith}. These additions push beyond pure gold reference matching, but exterior renders cannot resolve internal mating, and manifoldness may clear a 3D printer but not an engineering audit.

\paragraph{Part synthesis and assembly as disjoint problems.}
A second, more structural gap separates learned CAD generation from how human industry engineers actually work: prior work studies part synthesis and assembly disjointly. In real industrial engineering, much of the design effort goes into the joint itself, ensuring a part mates with its neighbors under tolerance, fits the bolt pattern of what it bolts to, clears the cable that runs past it, and carries the load that arrives through that interface; tolerance buildup across mates is itself a discipline with decades of literature~\citep{whitney2004mechanical,boothroyd2010product}. A jointable part is the hard output, not a free byproduct of having a closed manifold. Prior CAD generation works structurally skip this problem in one of two ways. The first group generates isolated parts and ignores assembly altogether, so the interfaces never have to align with anything~\citep{deepcad,cadsignet,cadllama,cadcoder,ttcq,cadrecode,cadrille,cadassistant,threedgpt}. The second keeps assembly in scope but starts from parts already extracted from working CAD assemblies (e.g., the Fusion 360 Gallery) that are jointable by construction~\citep{joinable,automate,fusion360gallery}. The model only has to predict joint axes or mate poses between known-fittable parts, never to author the mating geometry from scratch. Neither setting confronts the actual industrial design problem: producing a new part that must mate with specified neighbors and survive the loads that pass through that interface.

\subsection{Problem Statement: From Geometric Matching to Engineering Validation}
\label{sec:problem:gap}
\label{sec:problem:fea}

Industrial CAD design follows a workflow fundamentally different from how learned generators produce CAD. A human engineer iterates through a tight loop of authoring a dimensionally precise blueprint, rendering and walking around the part, taking section cuts, intuiting how it will respond under load, and revising. This workflow cannot translate directly to an LLM-driven loop, and three issues stand out. \textbf{Blueprint authoring.} LLMs cannot author blueprints with engineering-grade dimensional tolerance. Image generation models such as Nano Banana Pro can produce blueprint-like images~\citep{nanobananapro}, but they cannot deliver the dimensional precision required by drawing-to-CAD work~\citep{drawing2cad}. \textbf{Visual inspection.} LLMs cannot drive an iterative inspection loop. Computer use agents that drive a CAD viewer through screen control are too slow and noisy for a tight generation loop~\citep{anthropic_computer_use}. Real-time video encoders deliver pixel streams instead of measurements~\citep{videollm_online}, so the agent cannot read off the dimensions it would need to revise. \textbf{Physical validation.} Engineers run FEA, and so can the agent. The gap is that current CAD benchmarks rarely bind the output artifact to whether it can actually be built and used. Consider a representative prompt from our evaluation set.
\refstepcounter{briefex}\label{brief:bajaframe}
\begin{list}{}{\setlength{\leftmargin}{1em}\setlength{\rightmargin}{1em}}
\item[]
\itshape
\textbf{Brief~\thebriefex.}~Design a single-seat off-road tubular space frame from 25.4 mm OD by 3.05 mm wall 1018 DOM tubing, with primary and secondary members plus joint gussets, suspension pickup tabs, and engine mounts. The frame must survive 5/4/4/6 G impact, rollover, and 3.5 G hub bump, with a buckling load factor of at least 1.5.
\end{list}
A high IoU score on this design can still miss a pickup tab whose bolt pattern is off by a millimeter or a frame member that buckles at a quarter of its rated load. Worse, gold reference matching marks down any geometry that does not match the one curated reference, while a single specification admits many engineering-valid solutions. These benchmarks reward geometric resemblance, while engineering use requires parts that satisfy physical constraints. In this work, we use finite element analysis (FEA) as the engineering validation layer for this task. FEA predicts how a mechanical design responds to loading by discretizing the geometry into a mesh and solving for stresses, displacements, natural frequencies, and buckling load factors. These are the quantities an engineering audit asks for. We use CalculiX, a free, open-source three-dimensional FEA solver compatible with the Abaqus\footnote{Abaqus is a commercial FEA suite by Dassault Syst\`emes SIMULIA (\url{https://www.3ds.com/products/simulia/abaqus}).} input deck syntax~\citep{calculix}. To evaluate a candidate STEP file\footnote{STEP (ISO 10303, Standard for the Exchange of Product model data) is the ISO file format for exchanging 3D CAD models between systems. We target the AP242 application protocol used for mechanical assemblies (\url{https://www.iso.org/standard/66654.html}).}, the pipeline meshes the geometry via gmsh, splices the mesh and the candidate's named selectors with a spec-side analysis template, executes CalculiX, and parses the solver outputs against the declared requirement checks.

However, FEA on its own does not provide a benchmark. Each prompt must come paired with a structured set of pass/fail criteria the solver output can be checked against. To this end, we construct \textbf{\newbench{}}, a benchmark of 50 prompt-and-requirement pairs (20 single-part, 30 multi-part) drawn from a curated pool of 466 candidate briefs spanning patents and supplier datasheets, engineering standards (\href{https://standards.nasa.gov}{NASA-STD}, \href{https://ecss.nl}{ECSS}, \href{https://www.aisc.org}{AISC}, \href{https://quicksearch.dla.mil}{MIL-STD}, \href{https://www.fia.com/sport/regulations}{FIA Art.253}), regional industrial catalogs, and intercollegiate engineering competitions. Every brief is self-contained, with numeric limits written inline instead of being referenced from external standards, and every criterion is a parametric check the harness can evaluate without human interpretation. As a concrete example, Brief~\ref{brief:bajaframe} from \newbench{} expands into the six requirements of Table~\ref{tab:bajareqs}.

\begin{table}[t]
\centering
\caption{\footnotesize \textbf{Representative \newbench{} pass/fail criteria for Brief~\ref{brief:bajaframe}.} R1 checks tube sizing, R2 limits yielding, R3 bounds helmet-location deflection, and R4 checks rollover buckling margin. Each row reports the metric, threshold, applicable load cases, and whether the verdict comes from geometry or CalculiX.}
\label{tab:bajareqs}
\small
\begin{tabular}{l l l l l}
\toprule
ID & Metric & Threshold & Applies to & Source \\
\midrule
R1 & primary tube OD (mm) & $=$ 25.4 & design & geometric \\
R2 & max von Mises tube stress (MPa) & $\le$ 246.7 & LC1--LC4 & \texttt{*STATIC} \\
R3 & helmet location deflection (mm) & $\le$ 25 & LC1, LC4 & \texttt{*STATIC} \\
R4 & first mode load factor under LC4 & $\ge$ 1.5 & LC4 & \texttt{*BUCKLE} \\
\bottomrule
\end{tabular}
\end{table}

\section{Building a CAD-Agent Pipeline with Engineering Feedback}
\label{sec:method}

\subsection{Pipeline overview}
\label{sec:method:skills}

As CAD jobs given to AI systems become more constrained, tool dependent, and engineering facing, a single prompt-to-geometry call becomes a brittle abstraction. Our pipeline puts the LLM agent in charge of design decisions and uses a deterministic controller for execution, validation, and feedback routing~\citep{threedgpt,cadassistant,evocad,cadsmith}. The agent writes CadQuery, exports a STEP artifact, inspects feedback, and revises geometry and selector metadata, while the controller runs tools and returns compact reports for the next attempt. We use CadQuery Python as the agent's executable parametric CAD language with direct STEP export, following recent CAD-code generation work~\citep{cadcoder,ttcq,cadrecode,cadrille}.

At each attempt, the controller creates an isolated workspace and provides the same brief, deliverable contract, and tool bundle. The feedback tools are exposed as optional capabilities, so the agent decides whether to request planning, visual, or simulation feedback before submitting a revised artifact. The controller validates files, runs deterministic checks, parses requirement verdicts, and feeds concise reports into the next attempt.

\subsection{Blueprint skill for design planning}
\label{sec:method:blueprint}

For planning, the agent can use a blueprint skill to turn the engineering brief into an explicit design plan before writing CAD. The agent first writes a short design brief and a \texttt{blueprint.yaml} that records functional requirements, materials, load paths, interfaces, support and load selectors, and verification targets. The blueprint then decomposes each part into \emph{construction units}, where each unit is a small additive, subtractive, or modifier component drawn from a closed grammar of parametric primitives and modifier operations. Figure~\ref{fig:main} shows a representative sample. This gives the downstream CAD process three contracts. \textbf{Closed grammar} keeps the design inside auditable parametric primitives. \textbf{Envelope and interface locking} makes mating faces, split planes, hole patterns, and clearance regions explicit before CAD is written. \textbf{Acceptance claims} expose dimensional targets and functional assumptions to validators and retry feedback.

We package blueprinting as a model-agnostic skill so that different agent harnesses can use the same planning procedure. The full skill package spans 23 files, 1.5k lines, and nearly 50k characters, with planning advice, schema templates, release checklists, difficulty and quality rubrics, and reference modules covering scope, datums, geometry, interfaces, loads, materials, assembly, safety, and validation. In our experiments, compact CCX-specific versions of this skill are loaded by multiple models across Codex and Claude Code harnesses. During repair, FEA and rich-view findings are first encoded as blueprint changes, so geometry edits are made from an updated engineering plan. The full blueprint for Brief~\ref{brief:bajaframe} is in Appendix~\ref{app:fullblueprint}.

\subsection{Rich-view tool for visual revision}
\label{sec:method:richview}

\begin{figure*}[t]
\centering
\includegraphics[width=\textwidth]{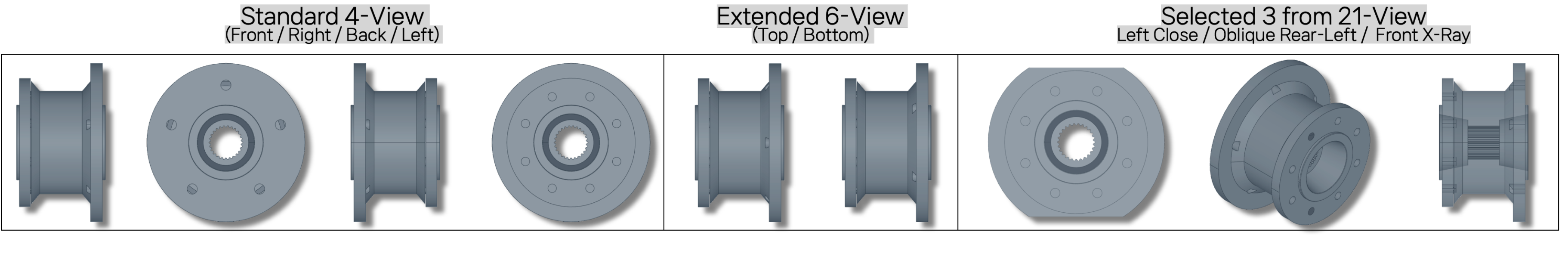}
\caption{Grouped nine-view sample for a generated wheel hub drawn from the 21-view rich-view set. The full set combines 12 axis-aligned and isometric views for exterior coverage, six close-ups for small features, and three alpha-blended x-ray views for internal mating and clearance. The strip contrasts conventional six-view coverage with selected additional views. The left close-up makes the bolt circle, concentric boss rings, and spline tooth profile auditable. The rear-left oblique ties rear-side reliefs and side cutouts to the hub depth. The front x-ray exposes the hidden axial stack.}
\label{fig:richviewwheelhub}
\end{figure*}

Once an initial CAD artifact exists, the agent can request a rich-view pass before final submission. The controller renders the assembled STEP through 21 calibrated ParaView\footnote{\url{https://www.paraview.org}} views and returns the image set as inspection context. This fixed coverage gives the agent static evidence similar to walking around the assembly, zooming into interfaces, and taking section cuts. Figure~\ref{fig:richviewwheelhub} shows a representative grouped subset, and Appendix~\ref{app:richview} lists the full view set.

The agent inspects the render set together with deterministic measurements of declared dimensions, mating expectations, and hole positions. The inspection prompt asks it to record compact typed fields including \texttt{verdict}, \texttt{summary}, \texttt{issues}, \texttt{failure\_category}, \texttt{primary\_claim\_id}, and \texttt{retry\_advice}. The report is small enough to fit a single agent context, and the views are broad enough that no external surface or internal mating face is hidden from inspection.

\subsection{Finite element analysis loop for engineering repair}
\label{sec:method:fea-feedback}

Once the agent finishes a submission, we run finite element analysis (FEA) on the submitted STEP artifact. In this study, the FEA step is placed \textbf{outside the agent} and executed by the controller. FEA may also be exposed as a free-use tool, but the controller-level placement makes one solver evaluation correspond to one feedback loop and keeps the test-time budget explicit. The controller evaluates the STEP artifact with the fixed \newbench{} CalculiX kit and writes a compact CCX feedback report. The report lists failed requirements, measured margins, selector or load-region problems, and analysis failures. The next attempt receives this report as engineering evidence while the canonical evaluation files, solver decks, and raw logs remain hidden.

Across repeated loops, the agent may see the target requirements and failed margins multiple times. While this may appear reminiscent of test leakage, it should be noted that it is natural that requirements are known from the start. This matches how engineers work with analysis feedback, optimizing towards a known requirement until the artifact satisfies all of them. The agent sees the generated CAD, notes, metadata, and summarized feedback, then revises the CadQuery program, selector metadata, blueprint when enabled, or design approximations before submitting a new STEP artifact. The same retry loop supports both the one-step FEA repair experiments in Section~\ref{sec:results:q1} and the longer test-time scaling runs in Section~\ref{sec:scaling}.

\section{Experimental setup}
\label{sec:setup}

\paragraph{Benchmarks.}
\label{sec:setup:benchmarks}
We evaluate on three benchmarks: \texttt{\newbench{}} (H-CCX), a sampled subset of \texttt{S2O} (Static-to-Openable)~\citep{s2o}, and a sampled subset of the \texttt{Fusion\,360 Gallery Assembly Dataset}~\citep{fusion360gallery}. \texttt{\newbench{}} (Section~\ref{sec:problem:fea}, 50 cases) is graded by the CalculiX harness via two metrics: \emph{Strict pass} counts items where every typed requirement passes, and \emph{Mean req pass} averages the per-case requirement pass fraction across the subset. Unlike \texttt{\newbench{}}, the two geometric benchmarks do not provide natural-language engineering prompts, so we generate them ourselves by querying GPT-5.4. Each call receives a rendered image of the target assembly together with structured metadata (part names, materials, counts, and articulation info). For \texttt{Fusion\,360}, these come from bundled renders and assembly manifests. For \texttt{S2O}, we rasterize the source mesh and use metadata from the dataset annotations. The model returns a multi-paragraph engineering description covering geometry, spatial relationships, inferred tolerances, material choices, articulation mechanics, and likely manufacturing process. We use this description as the evaluation prompt for all \texttt{S2O} and \texttt{Fusion\,360} experiments. Both datasets are restricted to the top 30\% of assemblies by face count to reduce compute while keeping the cases with the richest part counts and surface detail. This yields 133 cases for \texttt{S2O} and 225 cases for \texttt{Fusion\,360}. We score them with Chamfer distance (CD, $\downarrow$), F-score at $\tau{=}1\%$ ($\mathrm{F}_{1\%}$, $\uparrow$), and bounding-box IoU (Box-IoU, $\uparrow$).

\paragraph{Models.}
\label{sec:setup:models}
We run the main experiments through two production coding-agent harnesses, \texttt{Codex} and \texttt{Claude Code}. Codex is used for \texttt{GPT-5.5} and \texttt{GPT-5.4}, while Claude Code is used for \texttt{Opus-4.7} and \texttt{Sonnet-4.6}. We do not set custom generation parameters such as temperature, sampling settings, or token limits. Each run uses the default generation configuration chosen by its harness. For reasoning effort, we evaluate the highest and second-highest settings exposed by each harness. This is \texttt{xhigh} and \texttt{high} for GPT models in Codex, and \texttt{max} and \texttt{xhigh} for Claude Code. All runs receive the same task prompt and deliverable contract.

\section{Results}
\label{sec:results}

We organize the results around two questions: \textbf{Q1}: how do current frontier models perform under FEA-included evaluation (Section~\ref{sec:results:q1})? \textbf{Q2}: how do blueprint and image-based feedback change CAD generation (Section~\ref{sec:results:q2})?

\subsection{Q1: Frontier models under FEA-included evaluation}
\label{sec:results:q1}

Table~\ref{tab:modelperf} reports baseline performance on \newbench{} for the current frontier coding models from OpenAI Codex and Anthropic Claude Code, evaluated end-to-end through the CalculiX harness with no blueprint stage and no \imagefeedback{}.

\begin{table}[t]
\centering
\fontsize{8}{8.5}\selectfont
\caption{\footnotesize \textbf{Per-model performance on \newbench{} before and after one FEA-feedback repair round.} Cells report \textit{baseline}\,$\to$\,\textit{retry}. \colorbox{view21col}{Shading} marks improvement after feedback, with \textbf{bold} and \underline{underline} denoting best and second-best values within each column.}
\label{tab:modelperf}
\begin{tabular}{l l c c c c}
\toprule
& & \multicolumn{2}{c}{Single-part ($n{=}20$)} & \multicolumn{2}{c}{Multi-part ($n{=}30$)} \\
\cmidrule(lr){3-4} \cmidrule(lr){5-6}
Model & Reasoning & Strict pass & Mean req pass & Strict pass & Mean req pass \\
\midrule
\multicolumn{6}{l}{\textit{(a) Codex (OpenAI)}} \\
\midrule
\texttt{GPT-5.5} & \texttt{xhigh} & 0/20\,$\to$\,0/20 & \winretry{25.9\%\,$\to$\,\underline{39.4\%}}    & 0/30\,$\to$\,0/30                    & \winretry{\underline{15.0\%}\,$\to$\,\underline{36.1\%}} \\
\texttt{GPT-5.5} & \texttt{high}  & 0/20\,$\to$\,0/20 & \winretry{28.2\%\,$\to$\,38.1\%}                & \winretry{0/30\,$\to$\,\textbf{1/30}} & \winretry{8.3\%\,$\to$\,\textbf{39.2\%}}                 \\
\texttt{GPT-5.4} & \texttt{xhigh} & 0/20\,$\to$\,0/20 & \winretry{21.6\%\,$\to$\,38.0\%}                & 0/30\,$\to$\,0/30                    & \winretry{9.8\%\,$\to$\,26.8\%}                          \\
\texttt{GPT-5.4} & \texttt{high}  & 0/20\,$\to$\,0/20 & \winretry{23.3\%\,$\to$\,\textbf{42.1\%}}       & 0/30\,$\to$\,0/30                    & \winretry{8.4\%\,$\to$\,31.2\%}                          \\
\midrule
\multicolumn{6}{l}{\textit{(b) Claude Code (Anthropic)}} \\
\midrule
\texttt{Opus-4.7}   & \texttt{max}   & 0/20\,$\to$\,0/20 & \winretry{27.5\%\,$\to$\,29.1\%}                  & 0/30\,$\to$\,0/30 & \winretry{10.7\%\,$\to$\,18.6\%} \\
\texttt{Opus-4.7}   & \texttt{xhigh} & 0/20\,$\to$\,0/20 & \winbase{\textbf{32.7\%}\,$\to$\,31.4\%}          & 0/30\,$\to$\,0/30 & \winretry{6.6\%\,$\to$\,28.6\%}  \\
\texttt{Sonnet-4.6} & \texttt{max}   & 0/20\,$\to$\,0/20 & \winretry{25.4\%\,$\to$\,28.8\%}                  & 0/30\,$\to$\,0/30 & \winretry{6.2\%\,$\to$\,20.7\%}  \\
\texttt{Sonnet-4.6} & \texttt{xhigh} & 0/20\,$\to$\,0/20 & \winretry{\underline{28.7\%}\,$\to$\,33.3\%}      & 0/30\,$\to$\,0/30 & \winretry{11.8\%\,$\to$\,23.4\%} \\
\bottomrule
\end{tabular}
\end{table}

\paragraph{Current state-of-the-art agents fail to build engineering-sound products.}

Table~\ref{tab:modelperf} shows that \newbench{} is nearly unsolved under strict FEA grading. All models fail to build any strict-passing single-part or multi-part artifact except the Codex \texttt{GPT-5.5/high} FEA-retry run. Mean requirement pass gives a less binary view. On first attempt, Claude Code \texttt{Opus-4.7/xhigh} leads single-part generation at 32.7\%, outperforming the GPT counterparts in Table~\ref{tab:modelperf}. On multi-part generation, Codex \texttt{GPT-5.5/xhigh} has the strongest first-attempt score at 15.0\%, while \texttt{GPT-5.5/high} reaches the best post-retry score at 39.2\%. Multi-part cases add interface, clearance, load-transfer, and selector constraints beyond part geometry. \textbf{These results show that agents may satisfy some checks, but rarely produce complete artifacts that are engineering-usable.} This supports our earlier concern that evaluations based mainly on geometric similarity, validity, or visual plausibility can miss whether a CAD artifact satisfies a precise engineering contract.

\paragraph{Reasoning effort changes performance but not monotonically.}

Table~\ref{tab:modelperf} also shows that higher reasoning effort does not consistently improve performance. For example, in several Codex rows, \texttt{high} beats \texttt{xhigh} after retry. A similar pattern appears in Claude Code, where \texttt{Opus-4.7/xhigh} beats \texttt{Opus-4.7/max} on the first-attempt single-part score and on multi-part after retry. \textbf{This suggests that test-time scaling is not only about spending more compute, but about how the agent uses that compute.} We study this further in Section~\ref{sec:scaling}.

\paragraph{Some models are better first-shot designers, while others are better repair agents.}

Appendix Figure~\ref{fig:model-repair-scatter} separates initial design quality from the ability to use FEA feedback. These are not the same capability. On single-part tasks, Claude Code \texttt{Opus-4.7/xhigh} has the strongest first attempt at 32.7\%, but it slightly regresses after feedback. In contrast, Codex \texttt{GPT-5.4/high} starts lower at 23.3\% but gains 18.8 percentage points after one FEA-retry round. The same split appears in multi-part tasks. Codex \texttt{GPT-5.5/xhigh} has the strongest first-attempt GPT score at 15.0\%, but \texttt{GPT-5.5/high} produces the largest repair gain, improving by 30.9 points and reaching the best post-retry mean requirement pass. This suggests that first-shot CAD generation and feedback-driven repair should be treated as separate axes of model capability, rather than a single ranking.

\subsection{Q2: Evaluating feedback for CAD agents}
\label{sec:results:q2}

We test whether the typed blueprint stage (Section~\ref{sec:method:blueprint}) improves generation quality versus a baseline that emits CadQuery directly from the prompt. Two complementary experiments isolate the effect. Table~\ref{tab:q1q2} summarizes results for the blueprint stage (Panel~1) and \imagefeedback{} (Panel~2) across all benchmarks and view budgets.

\begin{table}[t]
  \centering
  \caption{Effect of blueprint planning and rich-view image feedback across \texttt{S2O}, \texttt{Fusion\,360}, and \newbench{}. Panel~1 compares direct generation against blueprint-guided generation; Panel~2 compares 0-view against 21-view feedback. Arrows show baseline\,$\to$\,after feedback, and \colorbox{view21col}{shading} marks cells where feedback improves the metric.}
  \label{tab:q1q2}
  \footnotesize
  \setlength{\tabcolsep}{1.1pt}
  \begin{tabular}{@{}l ccc ccc c@{}}
  \toprule
  & \multicolumn{3}{c}{S2O ($n{=}133$)} & \multicolumn{3}{c}{Fusion 360 ($n{=}225$)} & H-CCX ($n{=}50$) \\
  \cmidrule(lr){2-4} \cmidrule(lr){5-7} \cmidrule(lr){8-8}
  Model & Chamfer$\downarrow$ & F$_1$$\uparrow$ & IoU$\uparrow$ & Chamfer$\downarrow$ & F$_1$$\uparrow$ & IoU$\uparrow$ & \shortstack{Mean req\\pass$\uparrow$} \\
  \midrule
  \multicolumn{8}{l}{\textit{Panel 1: +blueprint only (Q1)}} \\
  \midrule
  GPT-5.5-xhigh
    & \wintwentyone{0.060\,$\to$\,0.047}
    & \wintwentyone{0.038\,$\to$\,0.045}
    & \wintwentyone{0.444\,$\to$\,0.592}
    & \wintwentyone{0.220\,$\to$\,0.163}
    & \wintwentyone{0.042\,$\to$\,0.067}
    & \wintwentyone{0.397\,$\to$\,0.482}
    & \winzero{0.194\,$\to$\,0.186} \\
  GPT-5.4-xhigh   
    & \wintwentyone{0.079\,$\to$\,0.044}
    & \wintwentyone{0.041\,$\to$\,0.049}
    & \wintwentyone{0.577\,$\to$\,0.608}
    & \wintwentyone{0.298\,$\to$\,0.169}
    & \wintwentyone{0.053\,$\to$\,0.063}
    & \wintwentyone{0.410\,$\to$\,0.505}
    & \wintwentyone{0.162\,$\to$\,0.175} \\
  Opus-4.7-Max
    & \wintwentyone{0.050\,$\to$\,0.048}
    & \wintwentyone{0.045\,$\to$\,0.047}
    & \wintwentyone{0.600\,$\to$\,0.614}
    & \winzero{0.152\,$\to$\,0.157}
    & \wintwentyone{0.055\,$\to$\,0.057}
    & \winzero{0.523\,$\to$\,0.508}
    & \wintwentyone{0.174\,$\to$\,0.203} \\
  Opus-4.7-xhigh 
    & \wintwentyone{0.048\,$\to$\,0.045}
    & \wintwentyone{0.044\,$\to$\,0.047}
    & {0.609\,$\to$\,0.608}
    & \wintwentyone{0.160\,$\to$\,0.159}
    & {0.063\,$\to$\,0.061}
    & \wintwentyone{0.494\,$\to$\,0.495}
    & \wintwentyone{0.170\,$\to$\,0.219} \\
  \midrule
  \multicolumn{8}{l}{\textit{Panel 2: +\imagefeedback{} only (Q2)}} \\
  \midrule
  GPT-5.5-xhigh
    & \winzero{0.060\,$\to$\,0.062}
    & \wintwentyone{0.038\,$\to$\,0.041}
    & \wintwentyone{0.444\,$\to$\,0.581}
    & \wintwentyone{0.220\,$\to$\,0.161}
    & \wintwentyone{0.042\,$\to$\,0.070}
    & \wintwentyone{0.397\,$\to$\,0.505}
    & \wintwentyone{0.194\,$\to$\,0.293} \\
  GPT-5.4-xhigh
    & \wintwentyone{0.079\,$\to$\,0.048}
    & \wintwentyone{0.041\,$\to$\,0.046}
    & \wintwentyone{0.577\,$\to$\,0.617}
    & \wintwentyone{0.298\,$\to$\,0.158}
    & \wintwentyone{0.053\,$\to$\,0.078}
    & \wintwentyone{0.410\,$\to$\,0.502}
    & \wintwentyone{0.162\,$\to$\,0.188} \\
  Opus-4.7-Max
    & \wintwentyone{0.050\,$\to$\,0.046}
    & \wintwentyone{0.045\,$\to$\,0.047}
    & \wintwentyone{0.600\,$\to$\,0.613}
    & {0.152\,$\to$\,0.153}
    & \wintwentyone{0.055\,$\to$\,0.061}
    & {0.523\,$\to$\,0.510}
    & \winzero{0.174\,$\to$\,0.171} \\
  Opus-4.7-xhigh
    & \winzero{0.048\,$\to$\,0.048}
    & \winzero{0.044\,$\to$\,0.044}
    & \winzero{0.609\,$\to$\,0.592}
    & \wintwentyone{0.160\,$\to$\,0.156}
    & {0.063\,$\to$\,0.059}
    & \wintwentyone{0.494\,$\to$\,0.503}
    & \wintwentyone{0.170\,$\to$\,0.207} \\
  \bottomrule
  \end{tabular}
  \end{table}

\paragraph{Feedback generally helps, but the degree of improvement varies.}
Table~\ref{tab:q1q2} shows that both feedback tools improve outputs in many settings, although the magnitude of improvement differs across tools, benchmarks, and models. The broad pattern matches Table~\ref{tab:modelperf}. Claude Code often starts from a stronger first attempt, while GPT-5 tends to gain more from feedback. Across the two GPT rows, both tools improve geometry on \texttt{S2O} and \texttt{Fusion\,360}, with average Box-IoU gains of about 0.09. On \newbench{}, \imagefeedback{} raises GPT mean requirement pass by 6.3 points on average, while blueprinting is nearly flat at 0.3 points. Claude Code shows smaller or mixed gains on geometry, but blueprinting gives a clearer \newbench{} gain, improving mean requirement pass by 3.9 points on average. These results support the same conclusion as Section~\ref{sec:results:q1}. First-attempt quality and feedback-driven improvement are related but distinct capabilities.

\section{Test-time scaling}
\label{sec:scaling}

Unlike recent findings~\citep{singh2025openai,agarwal2025gpt,hong2025kmmlu}, in Table~\ref{tab:modelperf}, we notice that expanded reasoning effort does not improve performance monotonically.  In contrast, asking the model to revise along with FEA feedback pushes the mean-requirement score to an average of 13.4 points across the Table~\ref{tab:modelperf} cells. This separates two forms of test-time compute. One spends more computation inside a single attempt. The other spends computation on a closed-loop interaction with an external evaluator. To study the second form, we fix the model and harness to Codex \texttt{GPT-5.5/high} and run all 50 \newbench{} cases for up to 10 attempts. The controller treats one FEA evaluation as one loop, returns typed checker verdicts and available visual reports to the agent, and records cumulative model time per item.

\paragraph{Repeated feedback scales when it becomes more concrete.}
Figure~\ref{fig:compute-scaling} shows the same broad but shallow first-retry pattern. Every plotted configuration improves from the first attempt to the second, yet higher reasoning effort does not consistently dominate lower-effort feedback runs. Codex \texttt{GPT-5.5/high} exceeds \texttt{GPT-5.5/xhigh}, and Claude Code \texttt{Opus-4.7/xhigh} exceeds \texttt{Opus-4.7/max}. The more important trend is that all settings continue to advance as retries compound. The two solid \texttt{GPT-5.5/high} curves isolate what is added to the loop. The lower solid curve repeats FEA retry without blueprinting. The higher and longer curve enables blueprinting, reaches better scores faster, and continues to improve through later attempts. The vertical markers show two feedback upgrades on this upper curve. At attempt 2, the agent receives the rich-view image review, which accounts for the first large jump. Until attempt 6, the FEA report mainly gives typed pass/fail verdicts. After attempt 6, detailed FEA feedback adds failure margins and identifies the relevant selector or load case, producing the second large jump. The final point reaches 60.5\% mean requirement pass with 9/50 strict-passing artifacts at 68 minutes per item. Overall, Figure~\ref{fig:compute-scaling} suggests that the useful compute is not just extra deliberation or extra planning. It is compute spent in a loop where the model receives increasingly concrete evidence about why the current CAD artifact fails.

\paragraph{Qualitative pass flips show what the loop repairs.}
\begin{wraptable}{r}{0.56\textwidth}
\vspace{-1.2em}
\centering
\caption{\footnotesize Strict-pass flips by repair type and whether conventional rendered-view feedback would likely expose the failure. Parenthesized letters mark the corresponding panels in Figure~\ref{fig:compute-scaling}.}
\label{tab:pass-flip-repairs}
\scriptsize
\setlength{\tabcolsep}{2.5pt}
\begin{tabular}{@{}p{0.36\linewidth} p{0.42\linewidth} p{0.16\linewidth}@{}}
\toprule
Case & Main repair & Visible? \\
\midrule
AISC column (C) & braced load path & partly \\
Prosthetic pylon & metric aliases and mass & no \\
RoboMaster launcher (B) & simpler mesh-stable body & partly \\
HPVC RPS & mass optimization & no \\
FIA rollcage & tube layout and compliance metadata & partly \\
UGV tool arm & hollow beam and selectors & partly \\
ECSS panel (D) & areal-density correction & no \\
AIJ structural & mesh-mass binding & no \\
KOSEN structural & mesh-mass aliases & no \\
\bottomrule
\end{tabular}
\vspace{-1.1em}
\end{wraptable}
We inspected the nine strict-passing artifacts from the longest \texttt{GPT-5.5/high} run using before/after renders, load overlays, field maps, and requirement-margin plots. The pass flips fall into four repair types. First, some retries perform real structural retuning. Figure~\ref{fig:compute-scaling}C is a clear example. The AISC steel column starts too slender and under-braced, failing compression, bending, and deflection checks, then passes after becoming a four-chord braced box column that redistributes stress while staying below the mass limit. The HPVC roll-protection system and UGV tool arm show similar shifts toward stiffer or lighter load-bearing structures. Second, some retries simplify mesh-sensitive geometry. Figure~\ref{fig:compute-scaling}B shows the RoboMaster launcher moving from a fragile firing structure with stress, displacement, and meshing failures to a simpler monolithic load path that lowers peak stress and radial growth. The FIA rollcage is similar, replacing an over-dense cage surrogate with a smaller tube layout and mount-compliance metadata. Third, some passes are checker-contract repairs. The prosthetic pylon, AIJ structural design, and KOSEN structural case mostly had the right physics, but failed because metric aliases, mass fields, or selector bindings were missing. Fourth, the ECSS spacecraft panel in Figure~\ref{fig:compute-scaling}D fixes a hidden mass-property error rather than a visible silhouette error, so a rendered-image judge would likely miss it. Its first version already has large stress, buckling, and modal margins but fails areal density by more than an order of magnitude; the passing retry preserves the panel planform while switching to a lightweight sandwich-equivalent representation, dropping projected areal density from about $70$ to $4.2\,\mathrm{kg/m^2}$.

\section{Conclusion}
\label{sec:conclusion}

In this work, we move CAD-agent evaluation toward engineering validation with finite-element requirements. We formulate generation from engineering briefs to assembled STEP artifacts and introduce \newbench{}, a 50-case benchmark with CalculiX evaluation kits and typed pass/fail checkers. We also study structured blueprints, rich-view inspection, and FEA feedback inside production agent harnesses. Together, these contributions show how current CAD agents behave when geometric construction, selector metadata, and physical requirements are evaluated as one engineering artifact. Quantitatively, one FEA-feedback round changes mean requirement pass by $-1.3$ to $+30.9$ points across Table~\ref{tab:modelperf} cells, with an average gain of 13.4 points, and the longest feedback loop raises \texttt{GPT-5.5/high} from 38.8\% to 60.5\% with 9/50 strict-passing artifacts. These results suggest that feedback design is an important direction for future CAD-agent research.


\newpage
\bibliographystyle{plainnat}
\bibliography{references}

\newpage
\appendix

\section{Additional model baseline results}
\label{app:modelperf-extra}

\begin{table}[h]
\centering
\caption{Additional OpenCode baseline results on \newbench{}. These rows use no FEA-retry round and are reported here to keep Table~\ref{tab:modelperf} focused on the main Codex and Claude Code comparison.}
\label{tab:modelperf-opencode}
\small
\begin{tabular}{l l c c c c}
\toprule
& & \multicolumn{2}{c}{Single-part ($n{=}20$)} & \multicolumn{2}{c}{Multi-part ($n{=}30$)} \\
\cmidrule(lr){3-4} \cmidrule(lr){5-6}
Model & Reasoning & Strict pass & Mean req pass & Strict pass & Mean req pass \\
\midrule
\texttt{GPT-5.5}         & \texttt{high} & 0/20 & 27.2\% & 1/30 & 13.9\% \\
\texttt{DeepSeek-V4-Pro} & \texttt{max}  & 0/20 & 28.6\% & 0/30 & 16.1\% \\
\texttt{Kimi-K2.6}       & \texttt{max}  & 0/20 & 21.8\% & 0/30 & 12.2\% \\
\texttt{GLM-5.1}         & \texttt{max}  & 0/20 & 15.9\% & 0/30 & 9.8\% \\
\bottomrule
\end{tabular}
\end{table}

The OpenCode rows fall in the same low partial-credit regime as the main production-agent first attempts: open-model baselines can satisfy a nontrivial fraction of typed requirements, but they still rarely produce complete engineering-valid artifacts. DeepSeek-V4-Pro, for example, reaches 28.6\% mean requirement pass on single-part cases and 16.1\% on multi-part cases, comparable to the first-attempt range in Table~\ref{tab:modelperf}. The \texttt{GPT-5.5/high} OpenCode row also differs from the \texttt{GPT-5.5/high} Codex row in Table~\ref{tab:modelperf}, especially on multi-part cases (13.9\% and 1/30 strict pass versus 8.3\% and 0/30 before retry). This suggests that the harness is part of the measured system: prompt wrapping, execution conventions, tool routing, and artifact handling can change the observed CAD-agent performance even when the underlying model family is similar.

\clearpage

\section{Full rich-view ablation results}
\label{app:q2-full-view-ablation}

\begin{table}[h]
\centering
\caption{Full \imagefeedback{} view-count ablation on \texttt{S2O}. Cells show \textit{0-view}\,$\to$\,\textit{7-view}\,$\to$\,\textit{21-view}. Cell shading marks the best nonzero view budget, with \colorbox{view0col}{7-view} and \colorbox{view21col}{21-view}. 0-view winners are unshaded.}
\label{tab:q2-full-view-ablation-s2o}
\small
\begin{tabular}{l ccc}
\toprule
Model & Chamfer$\downarrow$ & F$_1$$\uparrow$ & IoU$\uparrow$ \\
\midrule
GPT-5.5-xhigh
  & \winseven{0.060\,$\to$\,0.048\,$\to$\,0.062}
  & \winseven{0.038\,$\to$\,0.046\,$\to$\,0.041}
  & \winseven{0.444\,$\to$\,0.583\,$\to$\,0.581} \\
GPT-5.4-xhigh
  & \wintwentyone{0.079\,$\to$\,0.048\,$\to$\,0.048}
  & \wintwentyone{0.041\,$\to$\,0.041\,$\to$\,0.046}
  & \wintwentyone{0.577\,$\to$\,0.585\,$\to$\,0.617} \\
Opus-4.7-Max
  & \wintwentyone{0.050\,$\to$\,0.053\,$\to$\,0.046}
  & \wintwentyone{0.045\,$\to$\,0.044\,$\to$\,0.047}
  & \wintwentyone{0.600\,$\to$\,0.603\,$\to$\,0.613} \\
Opus-4.7-xhigh
  & {0.048\,$\to$\,0.049\,$\to$\,0.048}
  & \winseven{0.044\,$\to$\,0.045\,$\to$\,0.044}
  & \winzero{0.609\,$\to$\,0.590\,$\to$\,0.592} \\
\bottomrule
\end{tabular}
\end{table}

\begin{table}[h]
\centering
\caption{Full \imagefeedback{} view-count ablation on \texttt{Fusion\,360}. Cells show \textit{0-view}\,$\to$\,\textit{7-view}\,$\to$\,\textit{21-view}. Cell shading marks the best nonzero view budget, with \colorbox{view0col}{7-view} and \colorbox{view21col}{21-view}. 0-view winners are unshaded.}
\label{tab:q2-full-view-ablation-fusion}
\small
\begin{tabular}{l ccc}
\toprule
Model & Chamfer$\downarrow$ & F$_1$$\uparrow$ & IoU$\uparrow$ \\
\midrule
GPT-5.5-xhigh
  & \winseven{0.220\,$\to$\,0.153\,$\to$\,0.161}
  & \wintwentyone{0.042\,$\to$\,0.066\,$\to$\,0.070}
  & \winseven{0.397\,$\to$\,0.515\,$\to$\,0.505} \\
GPT-5.4-xhigh
  & \winseven{0.298\,$\to$\,0.158\,$\to$\,0.158}
  & \wintwentyone{0.053\,$\to$\,0.067\,$\to$\,0.078}
  & \winseven{0.410\,$\to$\,0.510\,$\to$\,0.502} \\
Opus-4.7-Max
  & {0.152\,$\to$\,0.161\,$\to$\,0.153}
  & \winseven{0.055\,$\to$\,0.064\,$\to$\,0.061}
  & \winseven{0.523\,$\to$\,0.529\,$\to$\,0.510} \\
Opus-4.7-xhigh
  & \winseven{0.160\,$\to$\,0.156\,$\to$\,0.156}
  & {0.063\,$\to$\,0.056\,$\to$\,0.059}
  & \winseven{0.494\,$\to$\,0.504\,$\to$\,0.503} \\
\bottomrule
\end{tabular}
\end{table}

Tables~\ref{tab:q2-full-view-ablation-s2o} and~\ref{tab:q2-full-view-ablation-fusion} show that more views are useful but not uniformly better. The 7-view budget sometimes matches or outperforms the 21-view budget, especially for simpler or exterior-dominated objects where additional close-up and x-ray views expose little new geometry. In those cases, extra images can add inspection burden without adding information. The 21-view budget remains valuable when small, occluded, or internal features matter, but the ablation suggests that the optimal visual context depends on the target geometry rather than only on view count.

\clearpage

\section{Sample S2O and Fusion 360 evaluation prompts}
\label{app:s2o-fusion-prompts}

\begin{figure}[t]
\centering
\begin{minipage}[t]{0.48\linewidth}
\centering
\includegraphics[width=0.86\linewidth]{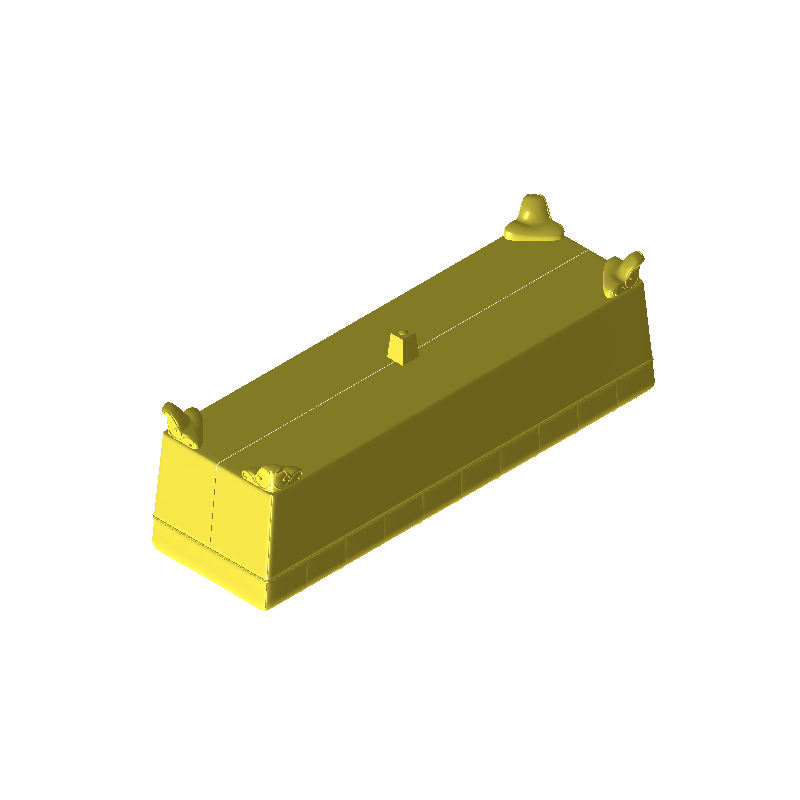}
\footnotesize
(a) \texttt{S2O}: storage bench, 70{,}102 faces.
\end{minipage}
\hfill
\begin{minipage}[t]{0.48\linewidth}
\centering
\includegraphics[width=0.86\linewidth]{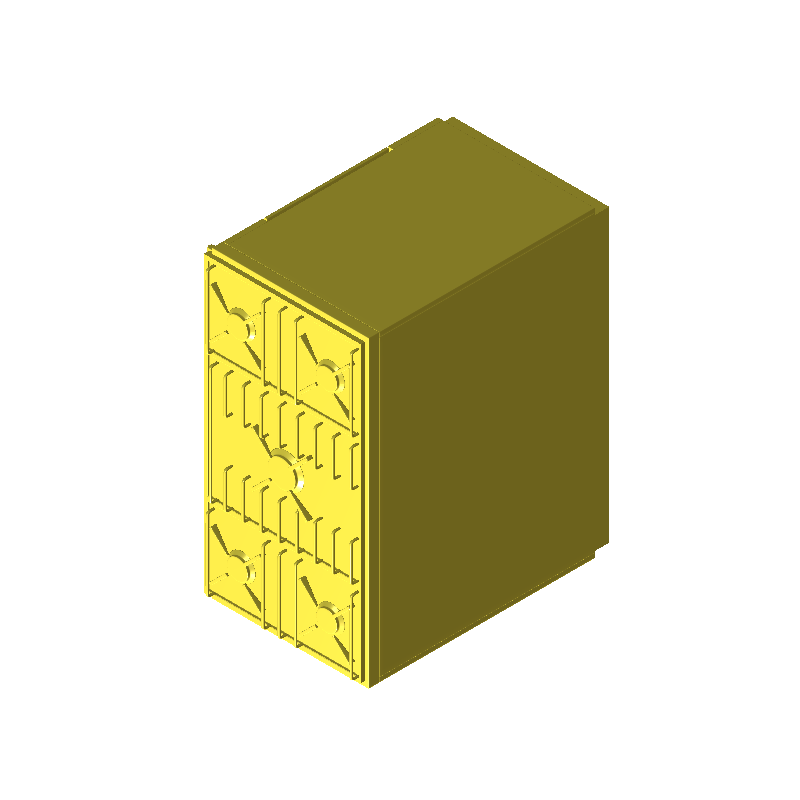}
\footnotesize
(b) \texttt{S2O}: kitchen appliance, 159{,}594 faces.
\end{minipage}
\caption{Representative \texttt{S2O} target items used to synthesize natural-language prompts.}
\label{fig:s2o-prompt-samples}
\end{figure}

\begin{figure}[t]
\centering

\begin{minipage}[t]{0.48\linewidth}
\centering
\includegraphics[width=0.86\linewidth]{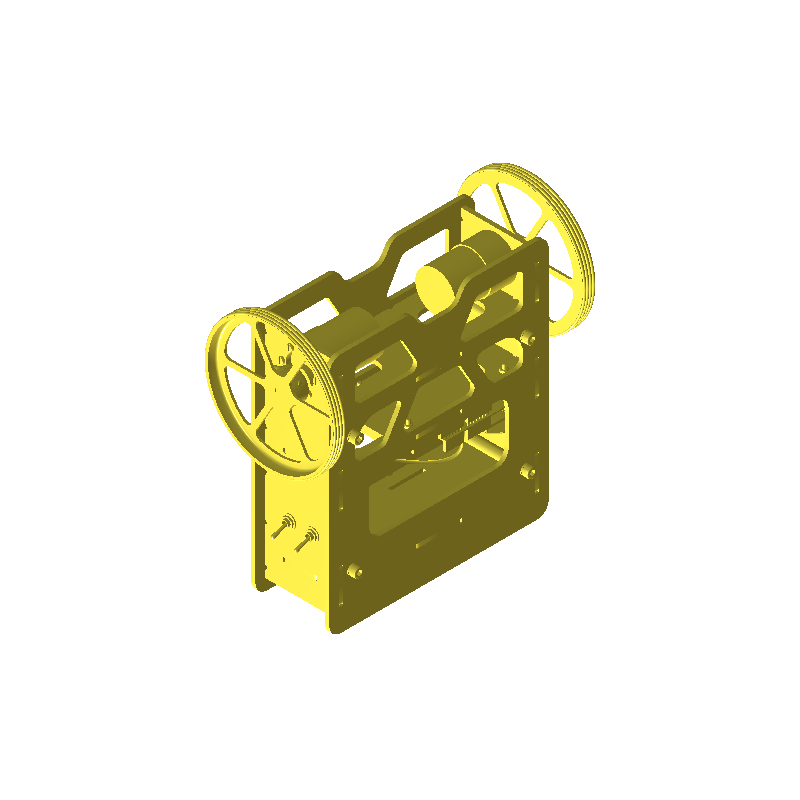}
\footnotesize
(c) \texttt{Fusion\,360}: robotic chassis, 5{,}789 faces.
\end{minipage}
\hfill
\begin{minipage}[t]{0.48\linewidth}
\centering
\includegraphics[width=0.86\linewidth]{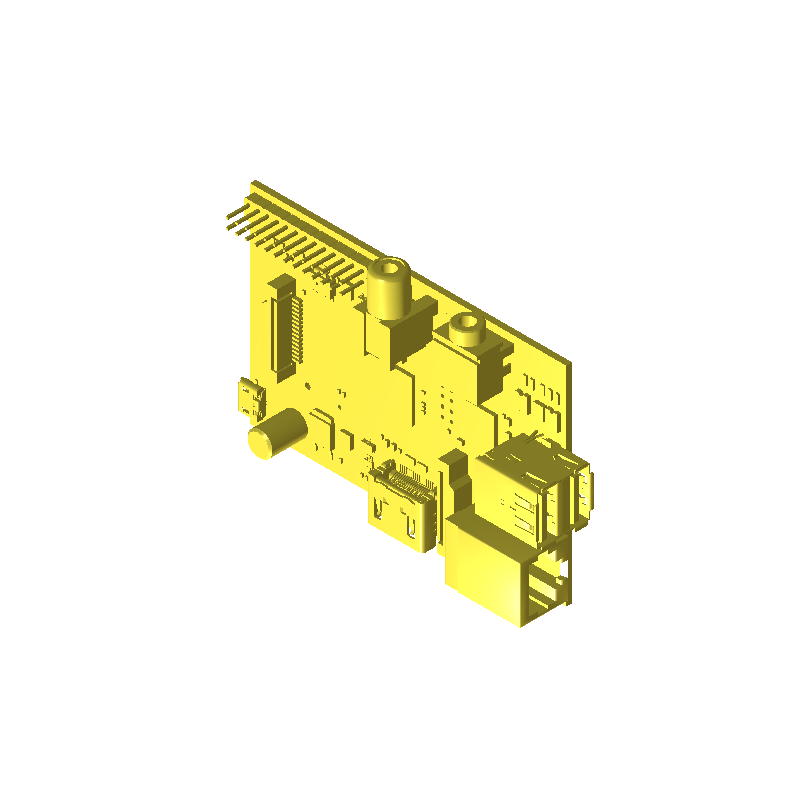}
\footnotesize
(d) \texttt{Fusion\,360}: I/O electronics board, 5{,}613 faces.
\end{minipage}
\caption{Representative \texttt{Fusion\,360} target items used to synthesize natural-language prompts.}
\label{fig:fusion360-prompt-samples}
\end{figure}

For the geometric benchmarks, each evaluation prompt is generated from the target rendering and structured metadata rather than written directly by the authors. Figures~\ref{fig:s2o-prompt-samples} and~\ref{fig:fusion360-prompt-samples} show representative target items, and the boxes below give the full corresponding natural-language prompts supplied to the CAD agent.

\subsection{S2O prompt samples}
\label{app:s2o-prompt-samples}

\begin{tcolorbox}[reqbox, title={Full generated prompt: \texttt{S2O} storage bench}]
The object is an elongated storage bench composed as a single box-form carcass with one top-opening lid. Geometrically, it reads as a long rectangular volume with straight side panels, closed ends, and a broad horizontal top surface. The proportions are consistent with bench furniture: longer in one direction, relatively narrow in depth, and low enough to serve as a seat while enclosing internal storage volume. The corners appear squared rather than heavily rounded, giving it a straightforward, utilitarian furniture profile.

Its only articulated feature is the lid, which likely rotates upward on a rear hinge running along the long back edge. With no visible drawers, doors, or segmented compartments indicated, the interior is probably accessed as one continuous storage cavity beneath the seat panel. This type of articulation is typical for blanket chests, entryway benches, or toy storage benches, where the top panel doubles as both seat and access cover. Depending on the design, the lid may include concealed hinges, a piano hinge, or soft-close support hardware to control motion and improve safety.

From a manufacturing standpoint, the bench is likely made from solid wood, plywood, MDF, or laminated particleboard assembled into a simple panel carcass. The top may be slightly thicker or reinforced to support seated loads, while the base panels provide torsional stiffness and enclosure. In use, it would fit in an entryway, bedroom, or living area, offering dual functionality as compact seating and hidden storage for shoes, linens, toys, or miscellaneous household items.
\end{tcolorbox}

\begin{tcolorbox}[reqbox, title={Full generated prompt: \texttt{S2O} kitchen appliance}]
The main structure would most likely be fabricated from stamped and folded sheet steel with an enamel or powder-coated finish, consistent with domestic cooking appliances. The drawer would ride on simple slide rails or formed sheet-metal guides, and the doors would use pin or concealed hinges attached to the front frame. Internally, the central chamber would be lined with heat-resistant metal and insulated from the outer shell. In use, the doors provide access to cooking or service compartments, while the drawer serves as a storage, broiler, or warming space. The design reads as a practical kitchen appliance with separated access zones and a durable, easy-to-clean construction.
\end{tcolorbox}

\clearpage

\subsection{Fusion 360 prompt samples}
\label{app:fusion360-prompt-samples}

\begin{tcolorbox}[reqbox, title={Full generated prompt: \texttt{Fusion\,360} robotic chassis}]
This is a fabricated steel electromechanical chassis consisting of numerous sheet-like and machined bodies assembled into a rigid open frame. The primary load-bearing members are two thick side plates with generous internal cutouts to reduce mass while preserving bending stiffness, tied together by a top cover and lower internal supports. Two large, lightweight spoked wheels with multi-rib rims are mounted on opposite sides of the frame at different vertical positions, indicating a drivetrain or balancing configuration rather than a conventional four-wheel cart. The wheel geometry suggests a relatively large diameter compared with the body width, likely chosen to improve obstacle traversal and provide rotational inertia. The top panel includes two toggle switches and elongated mounting slots, while the interior houses a vertically oriented PCB and a rectangular box-shaped component consistent with a battery, controller enclosure, or power electronics module.

Given the all-steel material assignment and the total mass of just over 5 kg, the structure is likely made from laser-cut or waterjet-cut sheet steel plates with bent or spacer-connected subassemblies, plus turned standoffs, fasteners, and wheel hubs. The repeated slots, access windows, and edge radii are characteristic of manufacturable plate design intended for easy assembly, wiring access, and weight reduction. The wheel spokes and rims may be machined, cut from plate, or fabricated as layered components. Fastener-mounted standoffs visible on the side plates imply modular internal mounting for electronics and serviceability. Tolerances for the frame plates are likely moderate, on the order of $\pm$0.1 to $\pm$0.3 mm for cut features, with tighter concentricity and fit requirements at wheel axle interfaces and bearing or hub locations.

In application terms, this assembly most likely serves as the chassis for a small robotic vehicle, self-balancing platform, or experimental mechatronic test rig. The open architecture supports rapid integration of control boards, wiring, and power components, while the robust steel construction favors durability over minimum weight. The design balances structural rigidity, accessibility, and manufacturability, and appears intended for prototype or low-volume production using standard sheet-metal fabrication, CNC-machined spacers, and mechanical fastening rather than cast or molded construction.
\end{tcolorbox}

\begin{tcolorbox}[reqbox, title={Full generated prompt: \texttt{Fusion\,360} electronics board}]
This is a single-board electronic I/O assembly consisting of a rectangular printed circuit board carrying multiple edge-accessible connectors, headers, and low-profile surface-mounted components. The board outline is simple and planar, with connectors concentrated on three sides for external access. A long through-hole pin header runs along the upper edge, indicating expansion or GPIO functionality. On the right-hand side, a large modular jack consistent with an RJ45 Ethernet connector is paired with a nearby rectangular shielded receptacle resembling a USB Type-A port. Near the lower center is another fine-pitch shielded connector that resembles HDMI or a similar high-density digital interface, while the lower left includes a small board-edge receptacle and a cylindrical side-facing jack consistent with a 3.5 mm audio or barrel-style connector. Two vertical cylindrical sockets near the upper half of the board suggest coaxial, antenna, or power-entry features. The remaining prismatic bodies likely represent integrated circuits, regulators, support components, and connector housings.

From the reported mass of about 0.14 kg and total volume of 17.9 cm$^3$, this is a dense but still hand-sized electronics module. The metadata lists all parts as steel, but in practical engineering terms the assembly would be a combination of FR-4 laminate PCB, copper circuitry, molded thermoplastic connector insulators, and stamped or die-cast metal shielding shells, with small amounts of solder. The larger connectors appear to be standard catalog components with formed sheet-metal shells and molded inserts, while the pin header would be a plated copper alloy leadframe in a plastic carrier. Typical tolerances would be driven by PCB fabrication and connector placement: board profile and hole locations likely within about $\pm$0.1 to $\pm$0.2 mm, with connector mating features controlled more tightly by the purchased component specifications.

Manufacturing would most likely involve standard PCB assembly processes: fabrication of the bare board, solder paste printing, automated placement of surface-mount parts, reflow soldering, followed by insertion and soldering of through-hole connectors such as the header, modular jack, and cylindrical jacks. The assembly context is consistent with a communications, embedded control, or single-board computing platform where multiple external interfaces are required on a compact board. The connector mix suggests data networking, peripheral connection, and possibly audio/video or power interfacing, making it likely part of a development board, industrial controller, or compact embedded gateway.
\end{tcolorbox}

\clearpage

\section{First-shot versus repair performance}
\label{app:repair-scatter}

\begin{figure}[h]
\centering
\includegraphics[width=\linewidth]{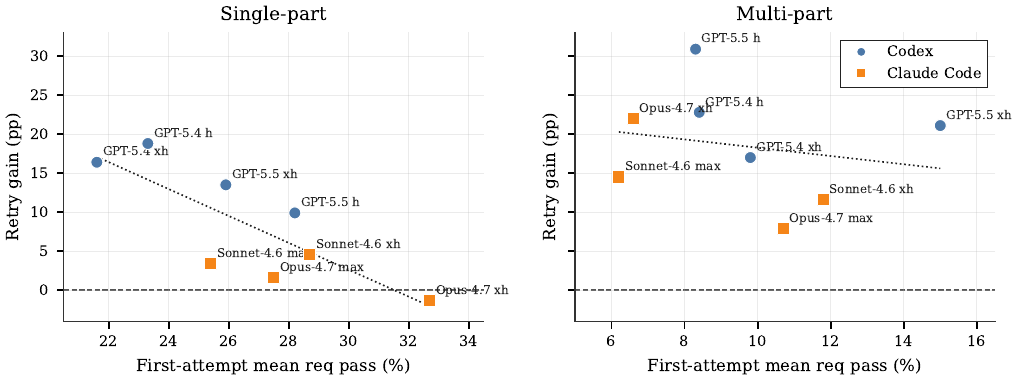}
\caption{First-attempt quality versus one-step FEA repair gain on \newbench{}. Each point is a Table~\ref{tab:modelperf} row with retry results. The x-axis reports first-attempt mean requirement pass. The y-axis reports the change in mean requirement pass, in percentage points, after one FEA-feedback round. Dotted lines show least-squares fits within each panel.}
\label{fig:model-repair-scatter}
\end{figure}

\clearpage

\section{Strict-pass retry case studies}
\label{app:strict-pass-retries}

\begin{figure}[h]
\centering
\setlength{\tabcolsep}{1.5pt}
\renewcommand{\arraystretch}{0.88}
\begin{tabular}{@{}p{0.20\linewidth}cccc@{}}
\toprule
& \footnotesize Surface fail & \footnotesize Surface pass & \footnotesize Field fail & \footnotesize Field pass \\
\midrule
\footnotesize\raggedright\textbf{AISC steel column}\newline Structural retuning &
\includegraphics[width=0.175\linewidth]{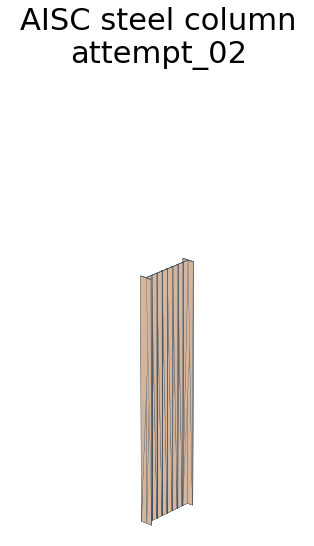} &
\includegraphics[width=0.175\linewidth]{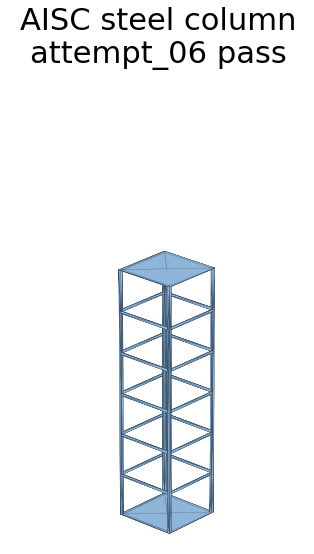} &
\includegraphics[width=0.175\linewidth]{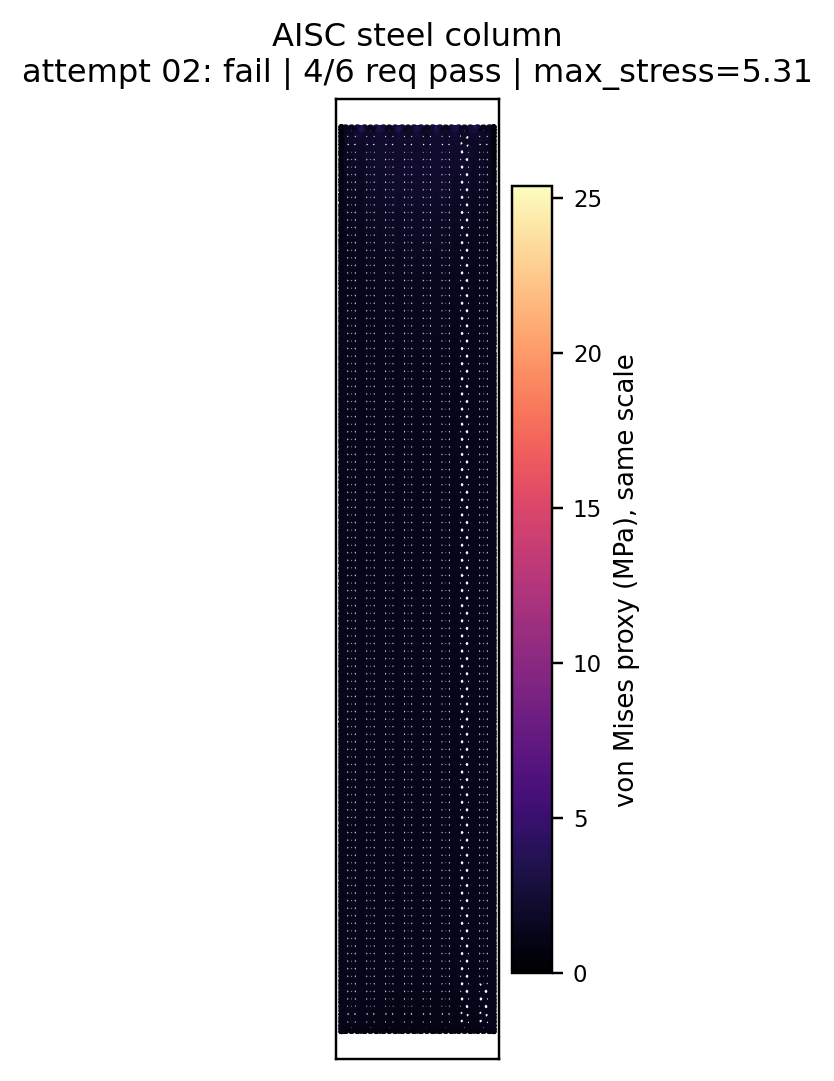} &
\includegraphics[width=0.175\linewidth]{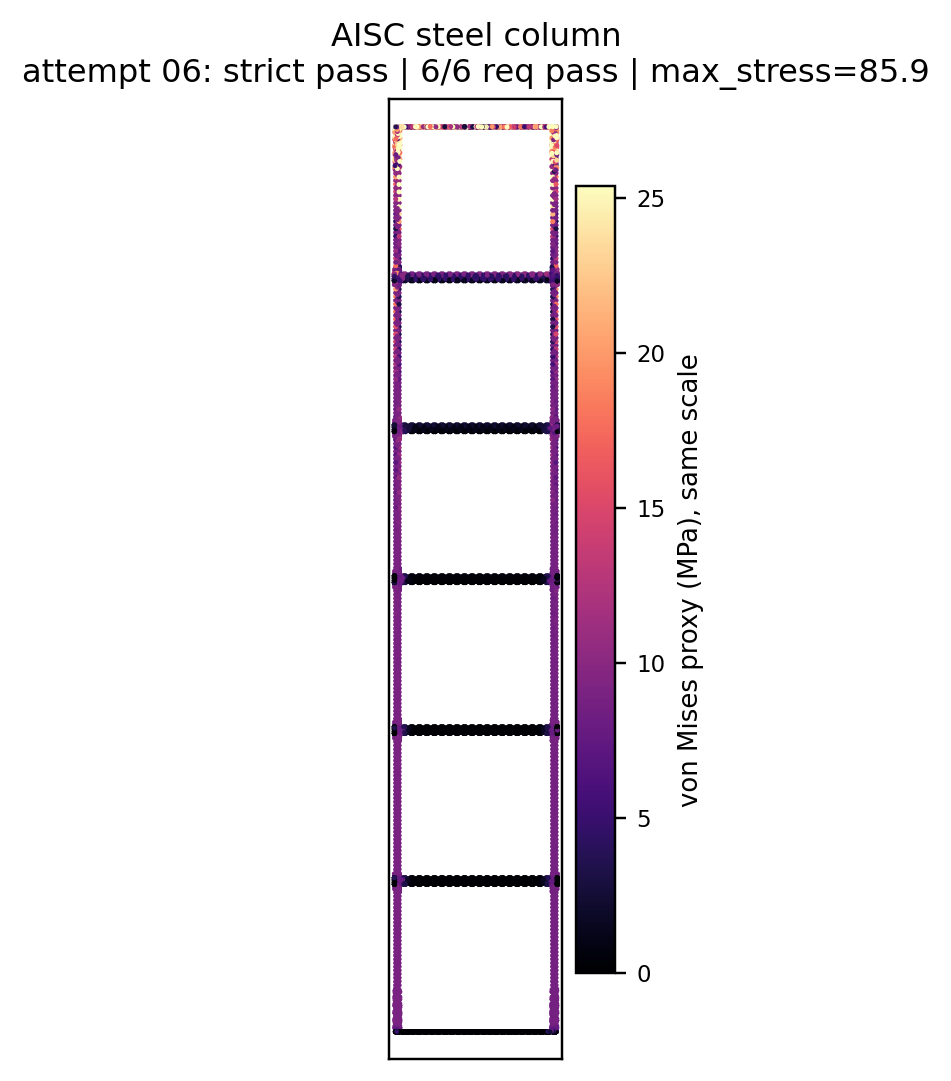} \\
\footnotesize\raggedright\textbf{HPVC roll-protection system}\newline Mass optimization &
\includegraphics[width=0.175\linewidth]{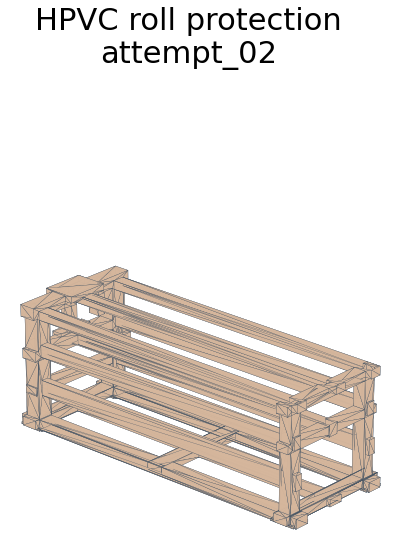} &
\includegraphics[width=0.175\linewidth]{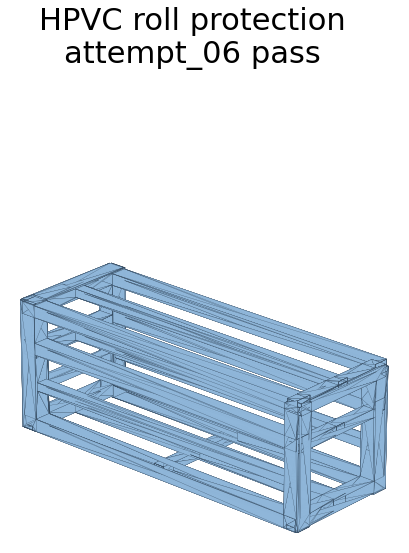} &
\includegraphics[width=0.175\linewidth]{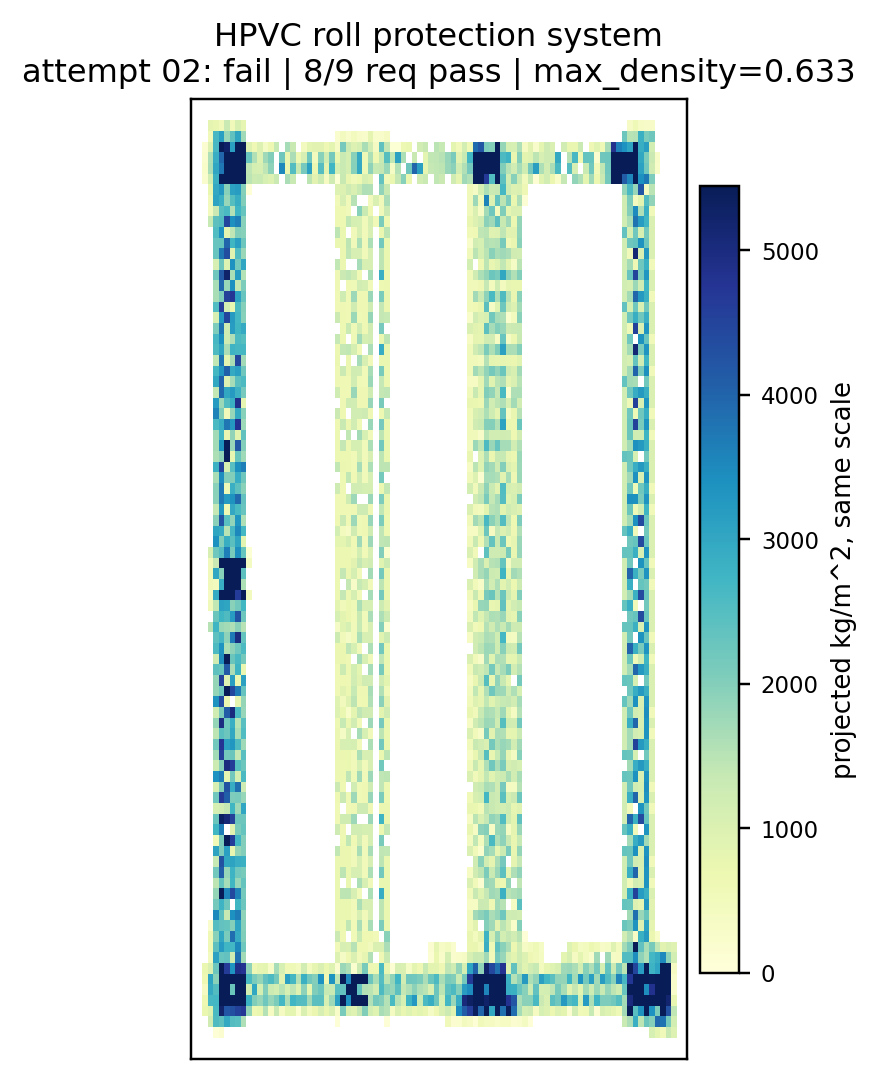} &
\includegraphics[width=0.175\linewidth]{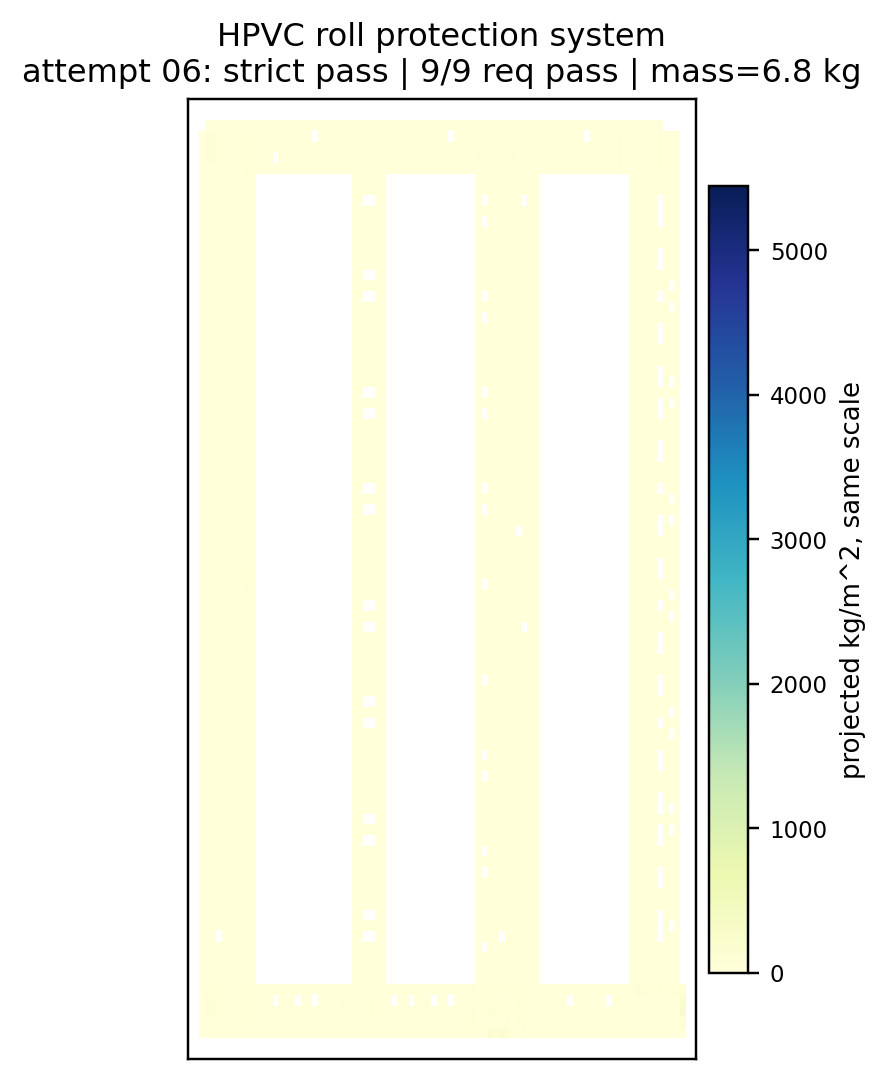} \\
\footnotesize\raggedright\textbf{UGV tool arm}\newline Hollow-beam retuning &
\includegraphics[width=0.175\linewidth]{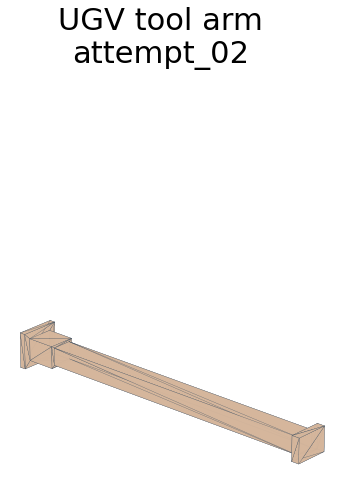} &
\includegraphics[width=0.175\linewidth]{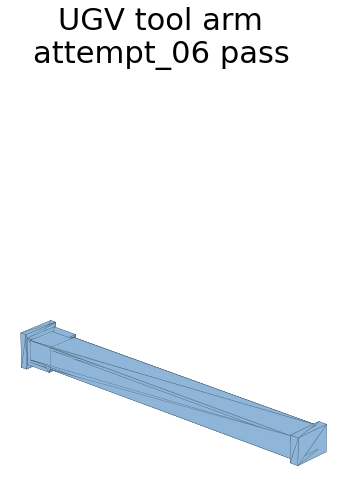} &
\includegraphics[width=0.175\linewidth]{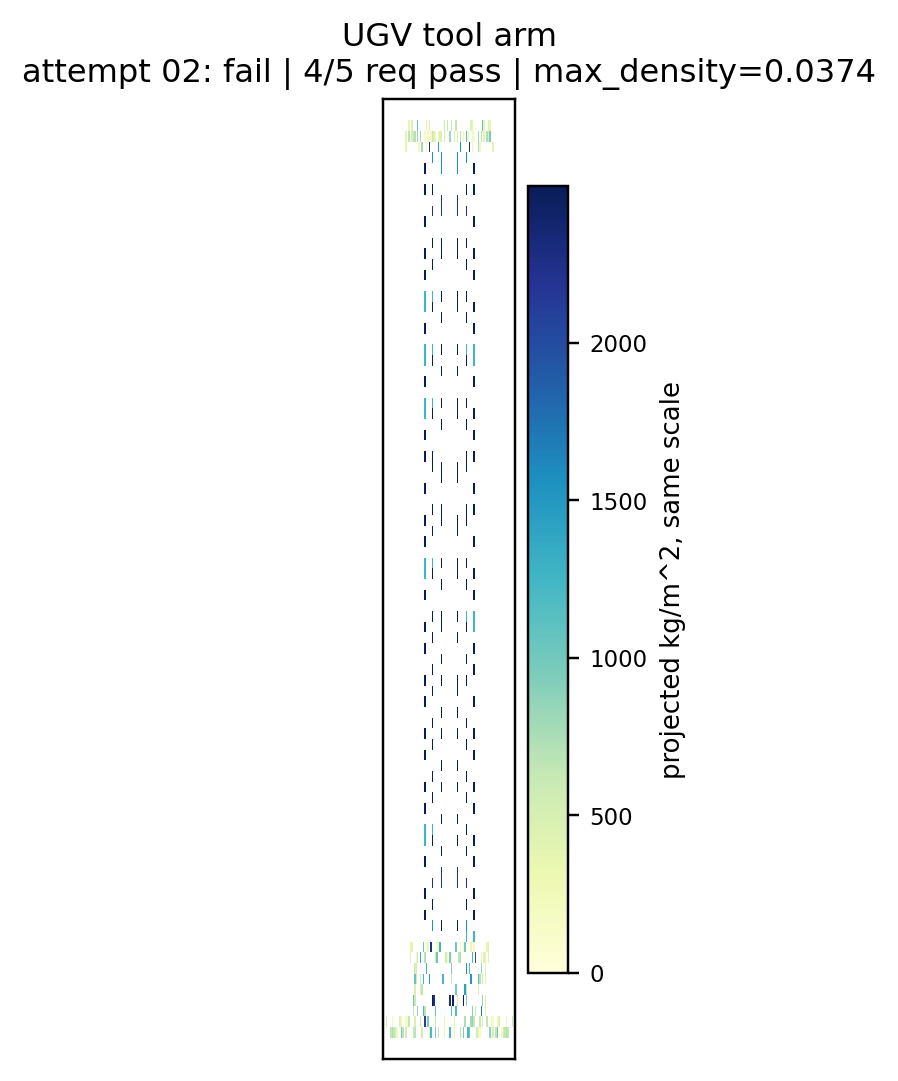} &
\includegraphics[width=0.175\linewidth]{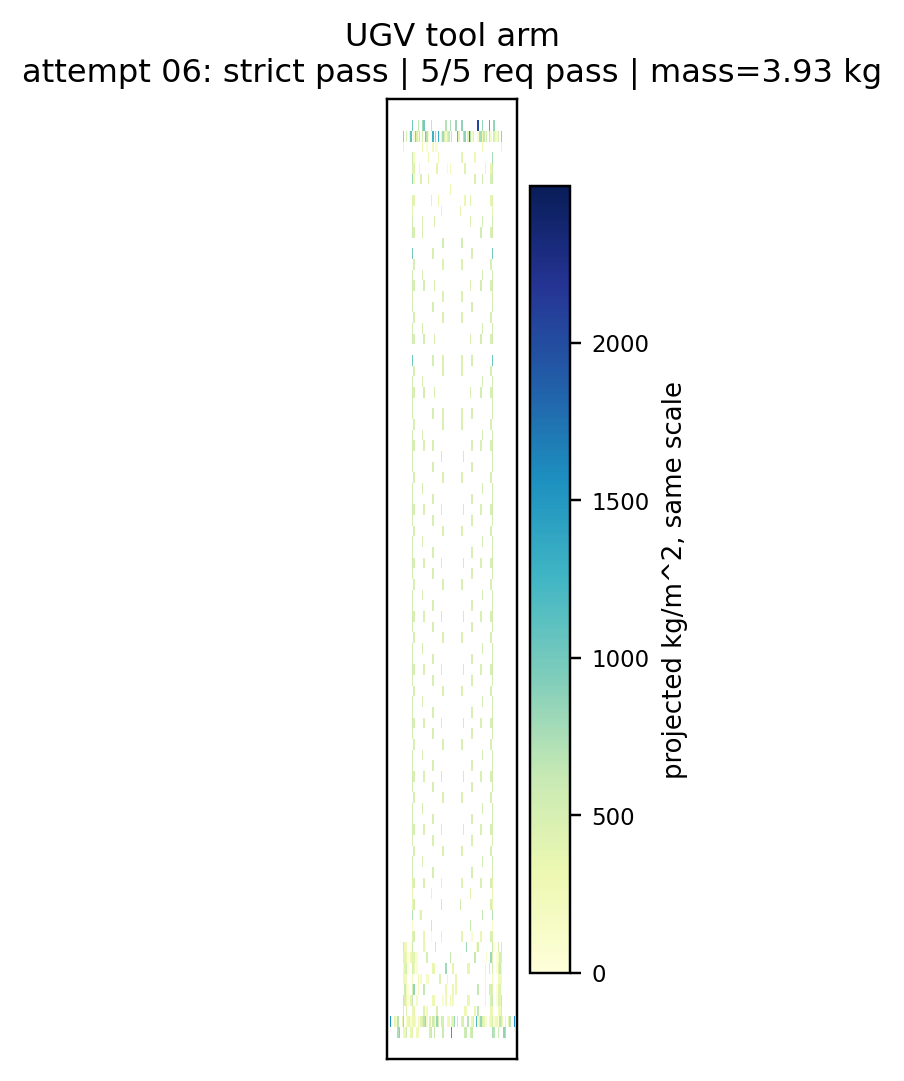} \\
\bottomrule
\end{tabular}
\caption{\footnotesize Strict-pass retries where the agent changes the physical load-bearing structure. The steel column becomes a braced four-chord box column, the HPVC roll-protection system sheds excess surrogate mass while preserving stiffness, and the UGV tool arm becomes a hollow box beam with cleaner root and tip selector faces.}
\label{fig:strict-pass-samples-structural}
\end{figure}

\begin{figure}[h]
\centering
\setlength{\tabcolsep}{1.5pt}
\renewcommand{\arraystretch}{0.88}
\begin{tabular}{@{}p{0.20\linewidth}cccc@{}}
\toprule
& \footnotesize Surface fail & \footnotesize Surface pass & \footnotesize Field fail & \footnotesize Field pass \\
\midrule
\footnotesize\raggedright\textbf{RoboMaster launcher}\newline Mesh-stable simplification &
\includegraphics[width=0.175\linewidth]{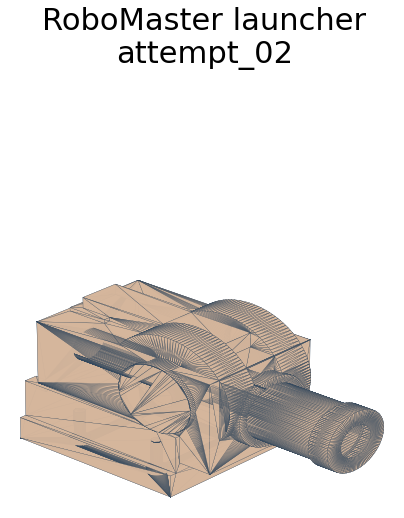} &
\includegraphics[width=0.175\linewidth]{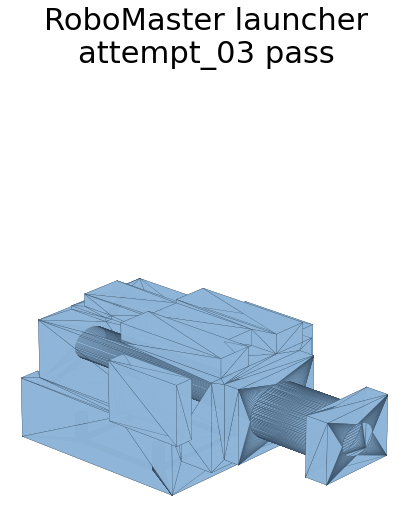} &
\includegraphics[width=0.175\linewidth]{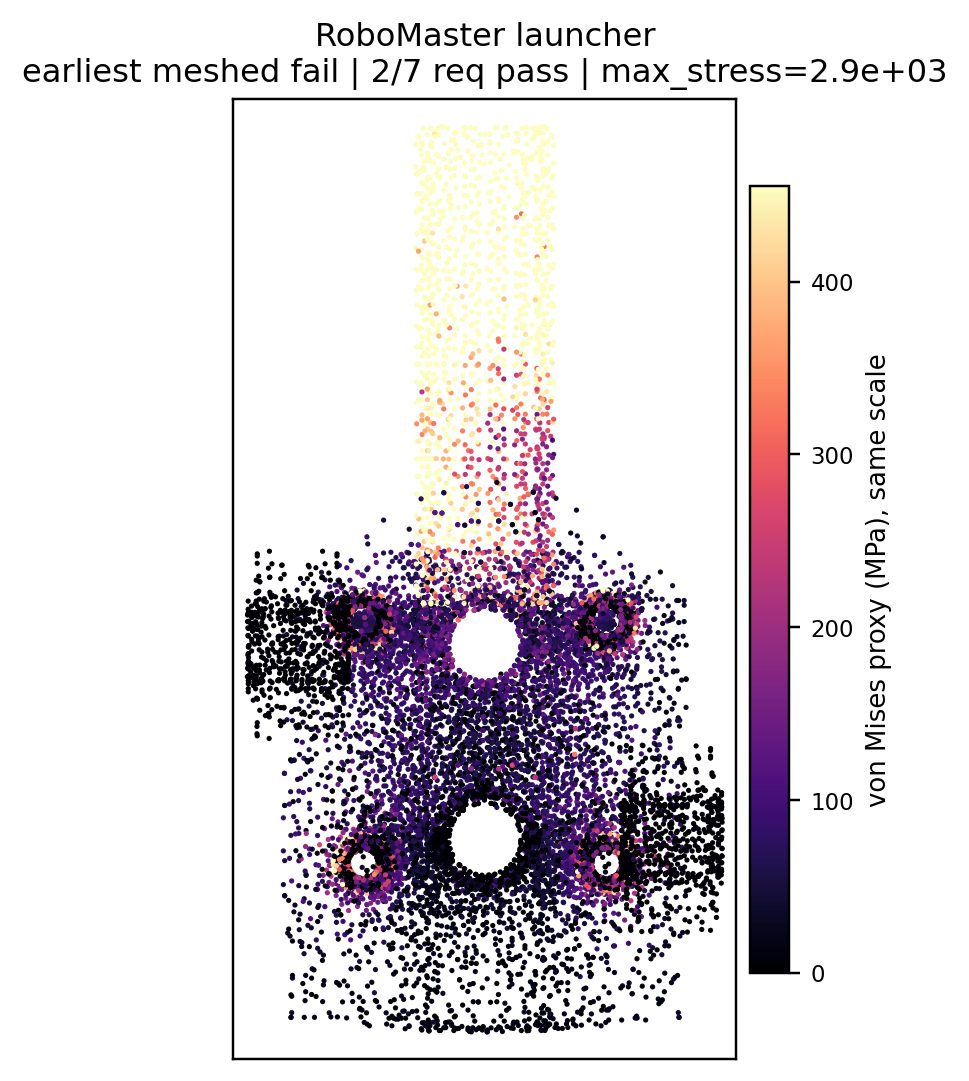} &
\includegraphics[width=0.175\linewidth]{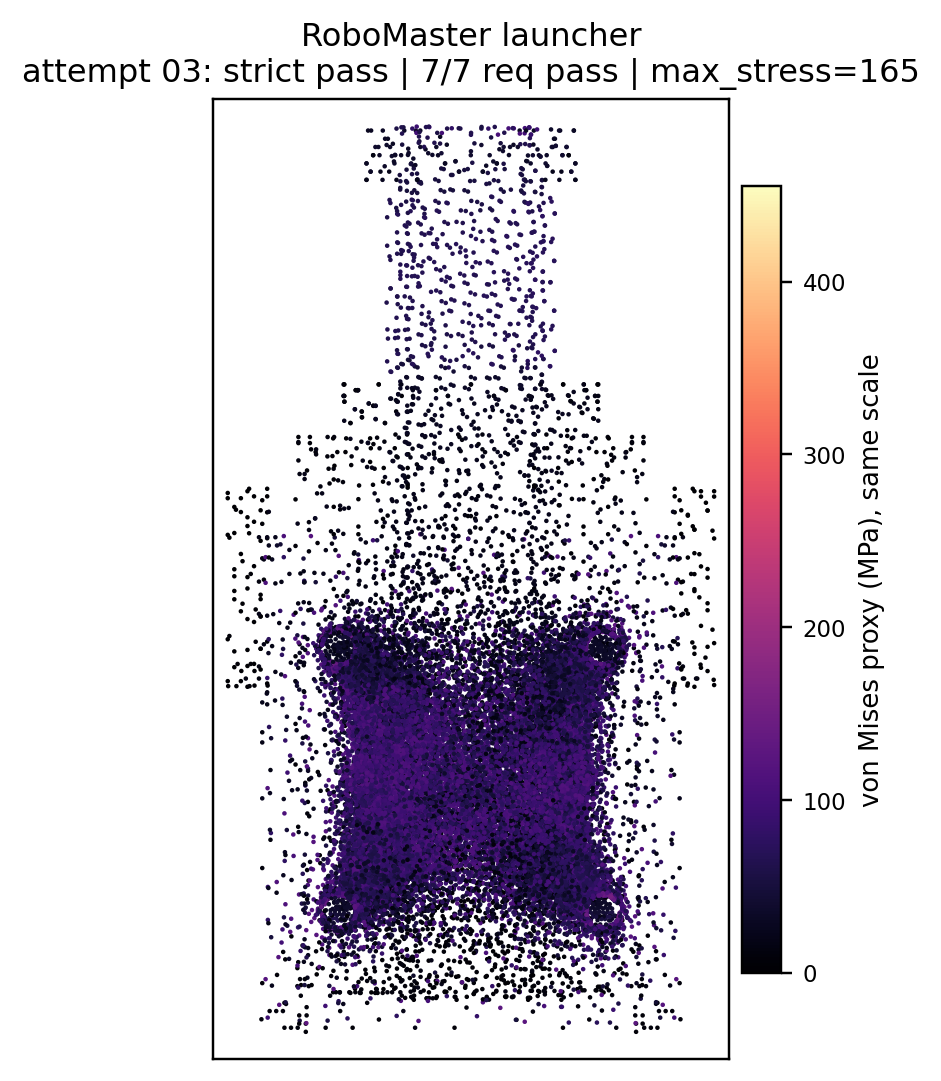} \\
\footnotesize\raggedright\textbf{FIA rollcage}\newline Tube-layout simplification &
\includegraphics[width=0.175\linewidth]{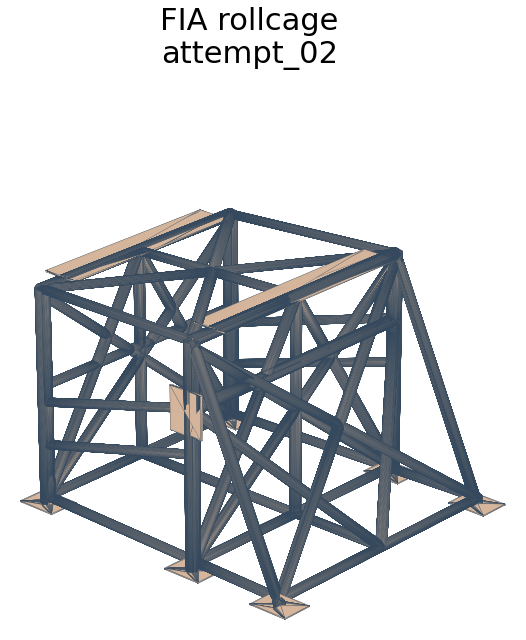} &
\includegraphics[width=0.175\linewidth]{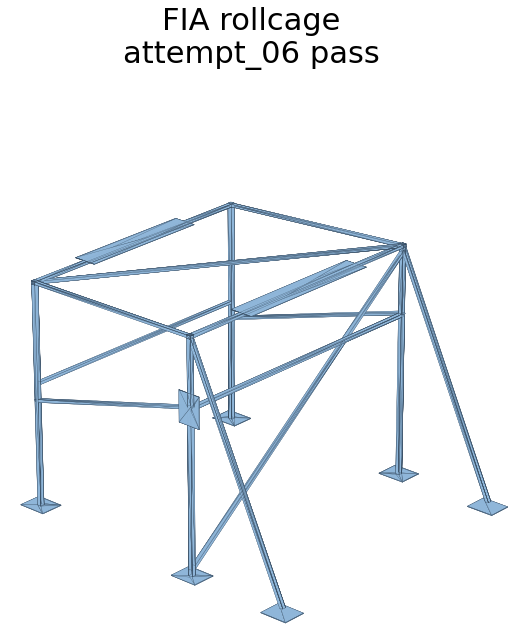} &
\includegraphics[width=0.175\linewidth]{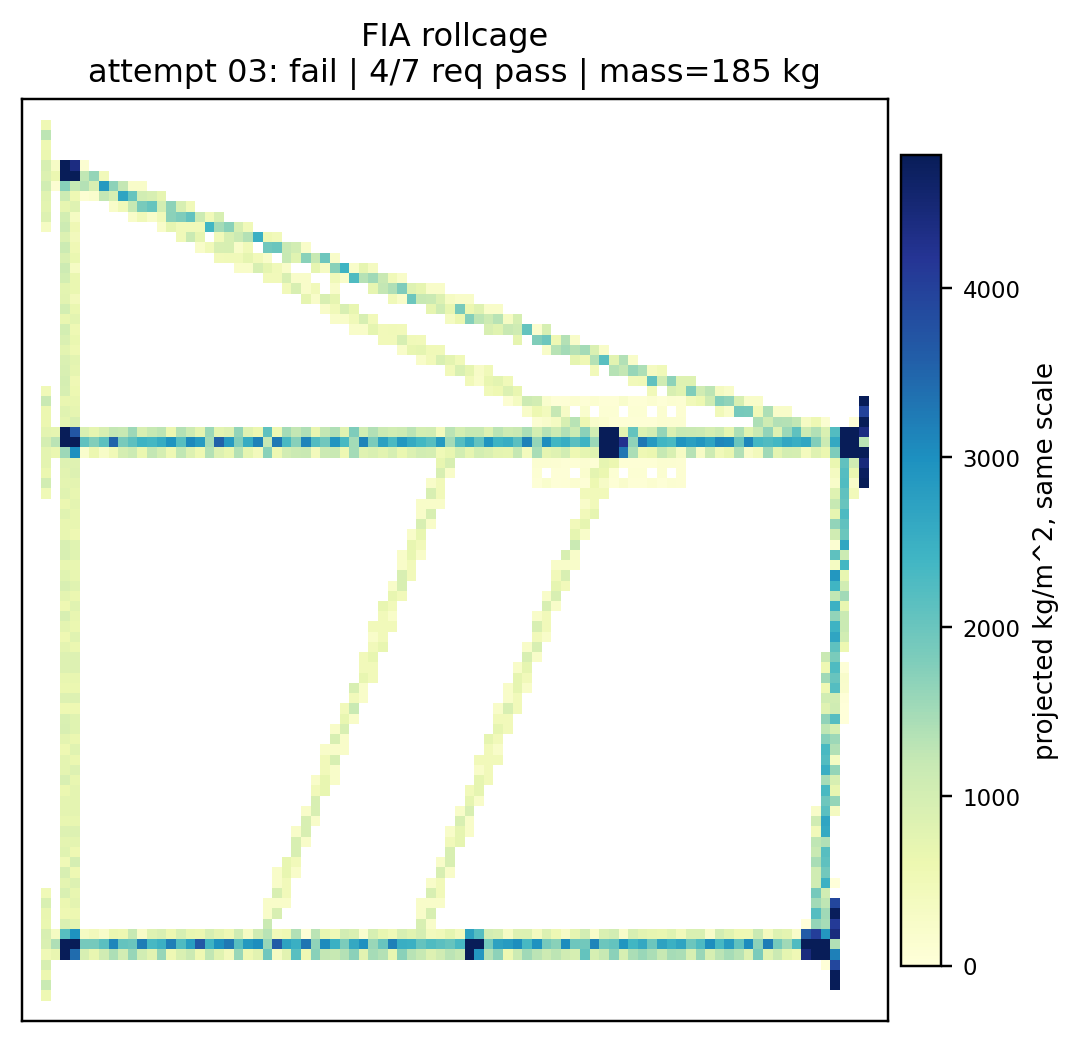} &
\includegraphics[width=0.175\linewidth]{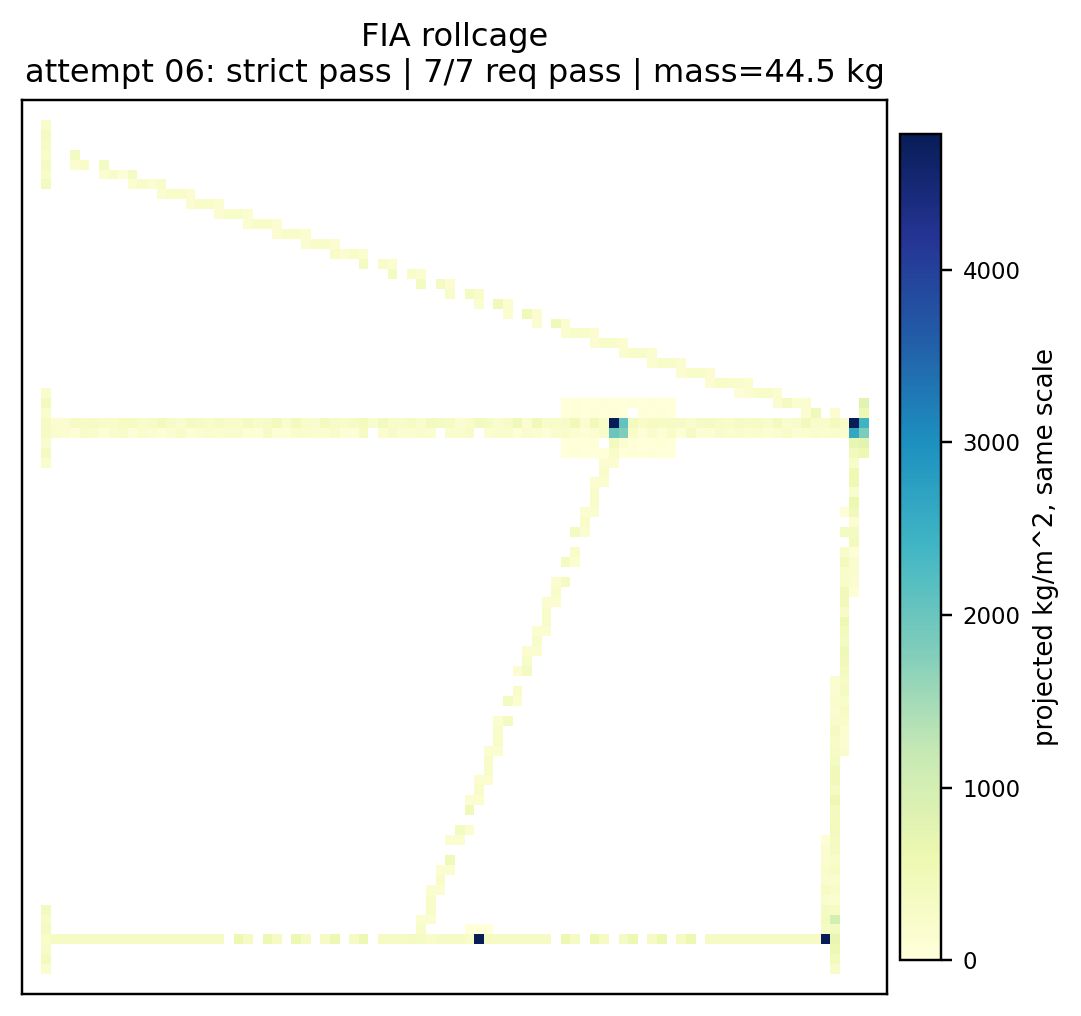} \\
\footnotesize\raggedright\textbf{ECSS spacecraft panel}\newline Areal-density correction &
\includegraphics[width=0.175\linewidth]{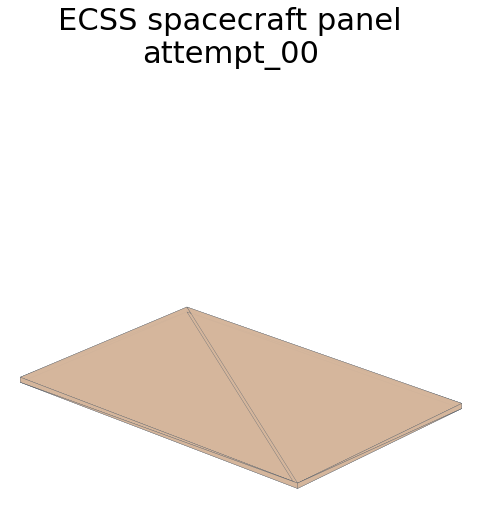} &
\includegraphics[width=0.175\linewidth]{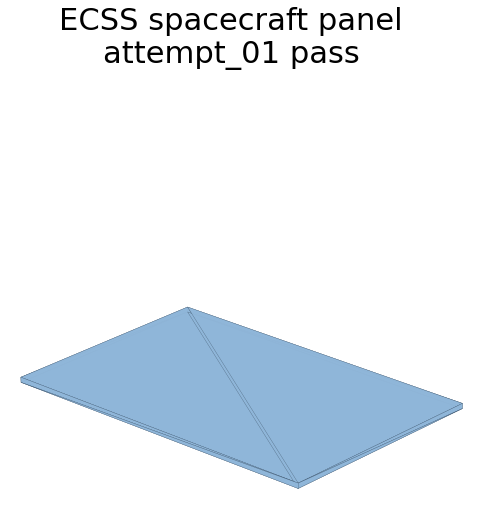} &
\includegraphics[width=0.175\linewidth]{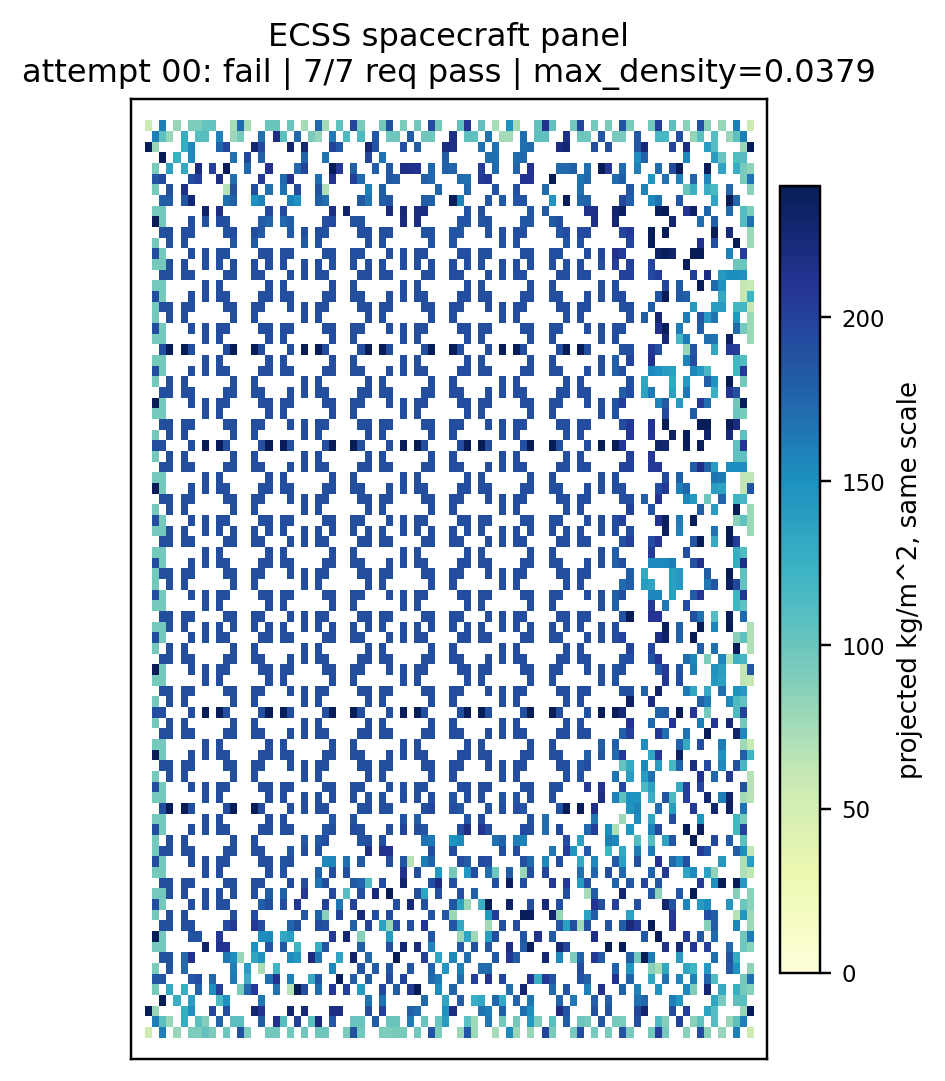} &
\includegraphics[width=0.175\linewidth]{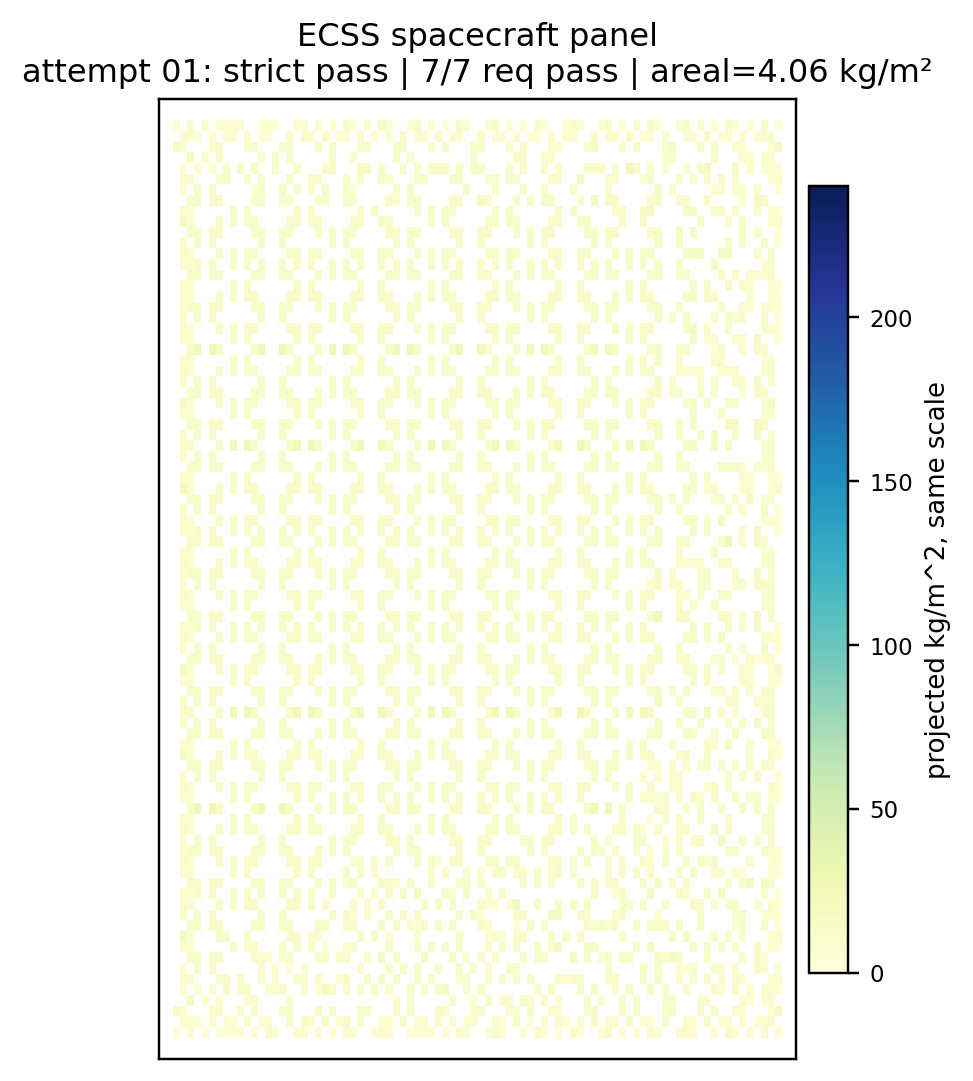} \\
\bottomrule
\end{tabular}
\caption{\footnotesize Strict-pass retries where the decisive change is simplification or hidden mass-property repair. The launcher removes fragile over-detailed geometry and routes load through a simpler body. The rollcage simplifies an unstable dense cage surrogate into an FIA-compatible tube layout with compliance metadata. The spacecraft panel looks similar in surface view, but the field map reveals the density correction that makes the panel pass.}
\label{fig:strict-pass-samples-simplification}
\end{figure}

\begin{figure}[h]
\centering
\setlength{\tabcolsep}{1.5pt}
\renewcommand{\arraystretch}{0.88}
\begin{tabular}{@{}p{0.20\linewidth}cccc@{}}
\toprule
& \footnotesize Surface fail & \footnotesize Surface pass & \footnotesize Field fail & \footnotesize Field pass \\
\midrule
\footnotesize\raggedright\textbf{ISO prosthetic pylon}\newline Metric aliases and mass &
\includegraphics[width=0.175\linewidth]{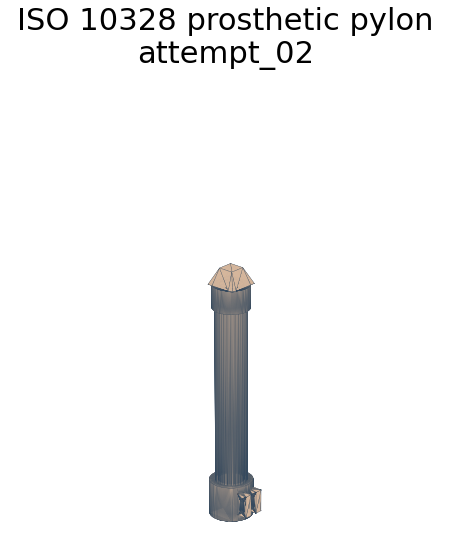} &
\includegraphics[width=0.175\linewidth]{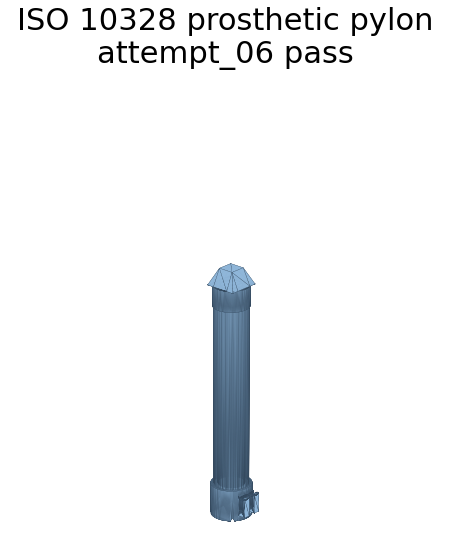} &
\includegraphics[width=0.175\linewidth]{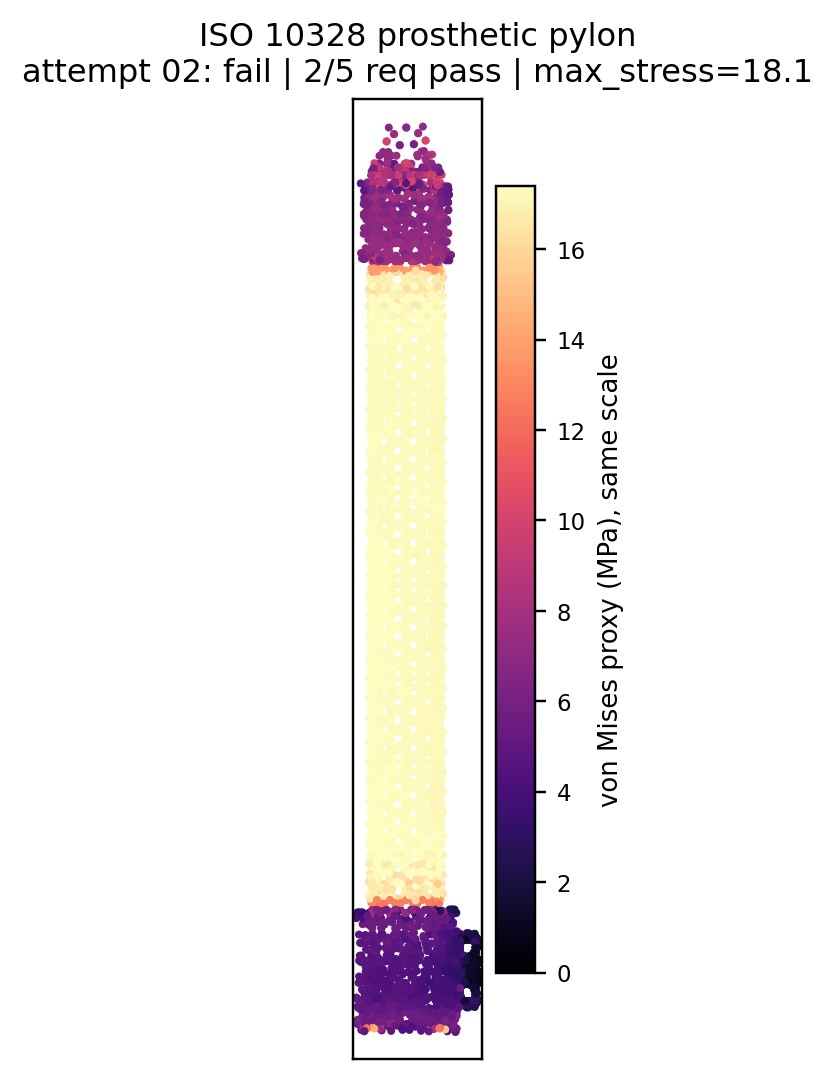} &
\includegraphics[width=0.175\linewidth]{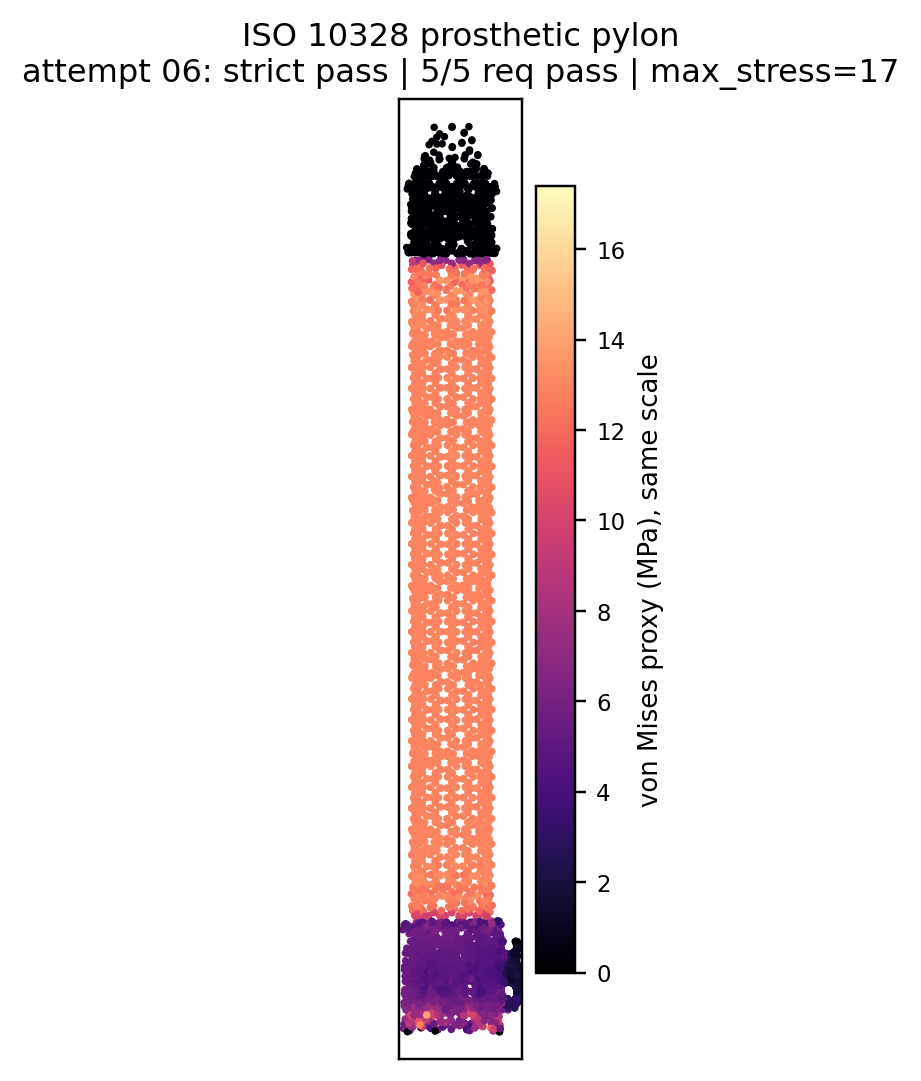} \\
\footnotesize\raggedright\textbf{AIJ structural design}\newline Mesh-mass binding &
\includegraphics[width=0.175\linewidth]{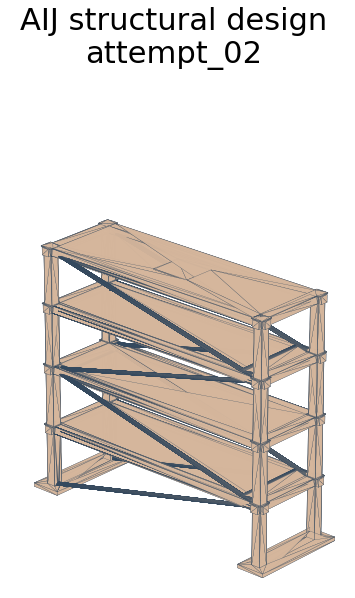} &
\includegraphics[width=0.175\linewidth]{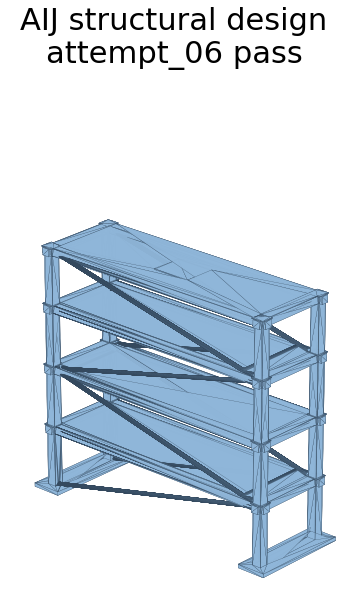} &
\includegraphics[width=0.175\linewidth]{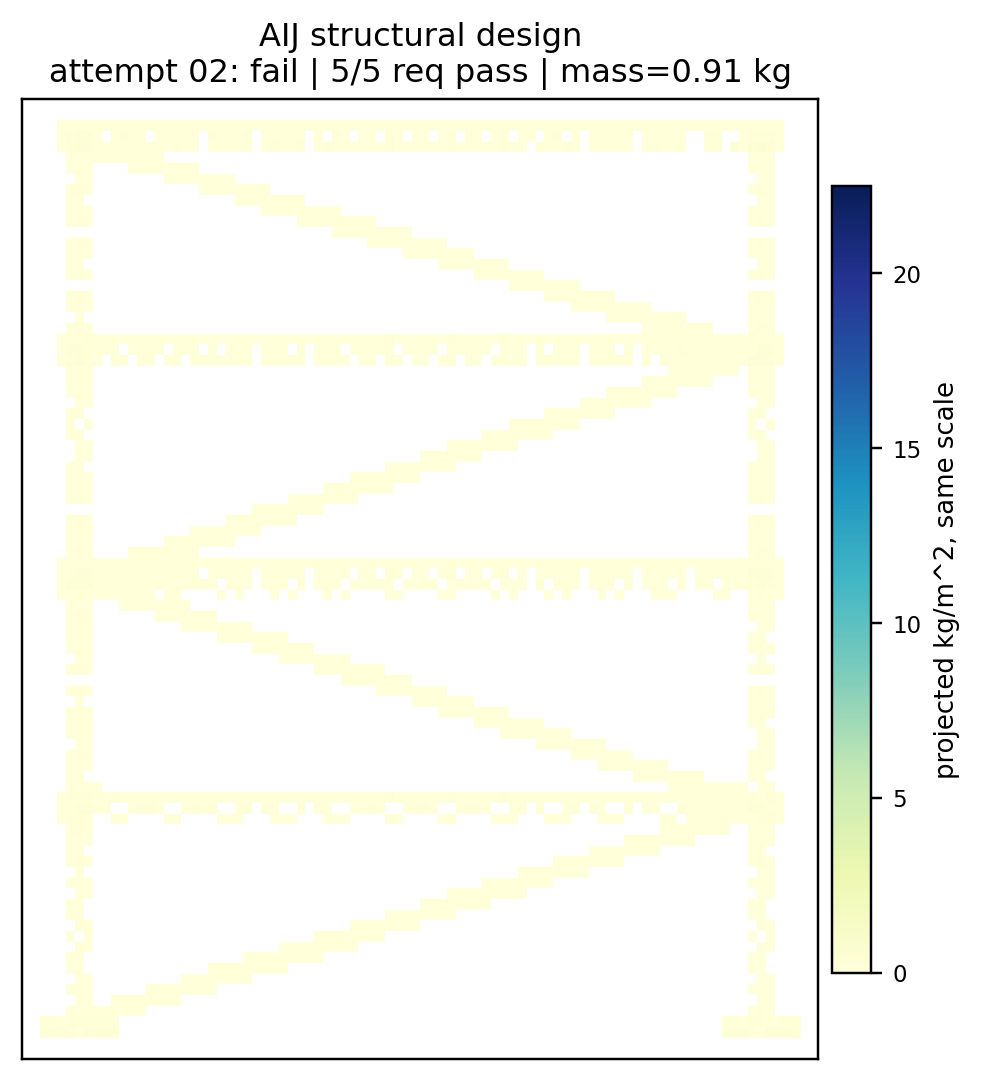} &
\includegraphics[width=0.175\linewidth]{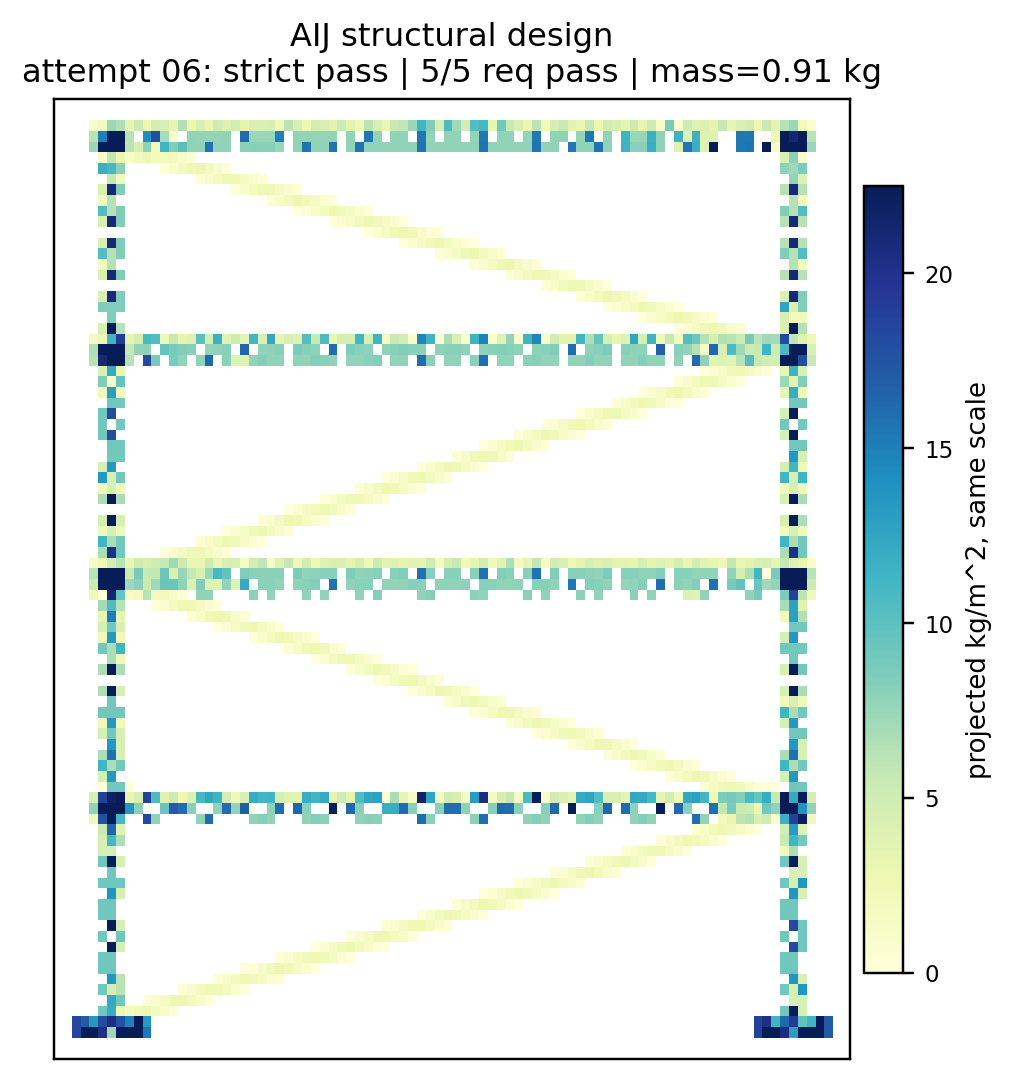} \\
\footnotesize\raggedright\textbf{KOSEN structural design}\newline Mass aliases &
\includegraphics[width=0.175\linewidth]{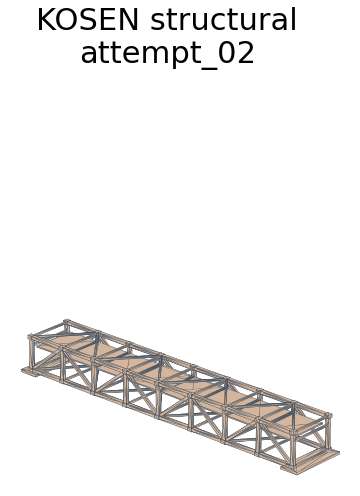} &
\includegraphics[width=0.175\linewidth]{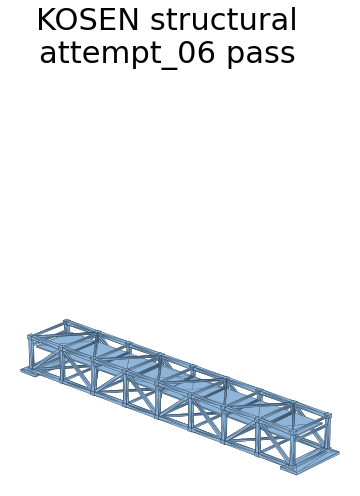} &
\includegraphics[width=0.175\linewidth]{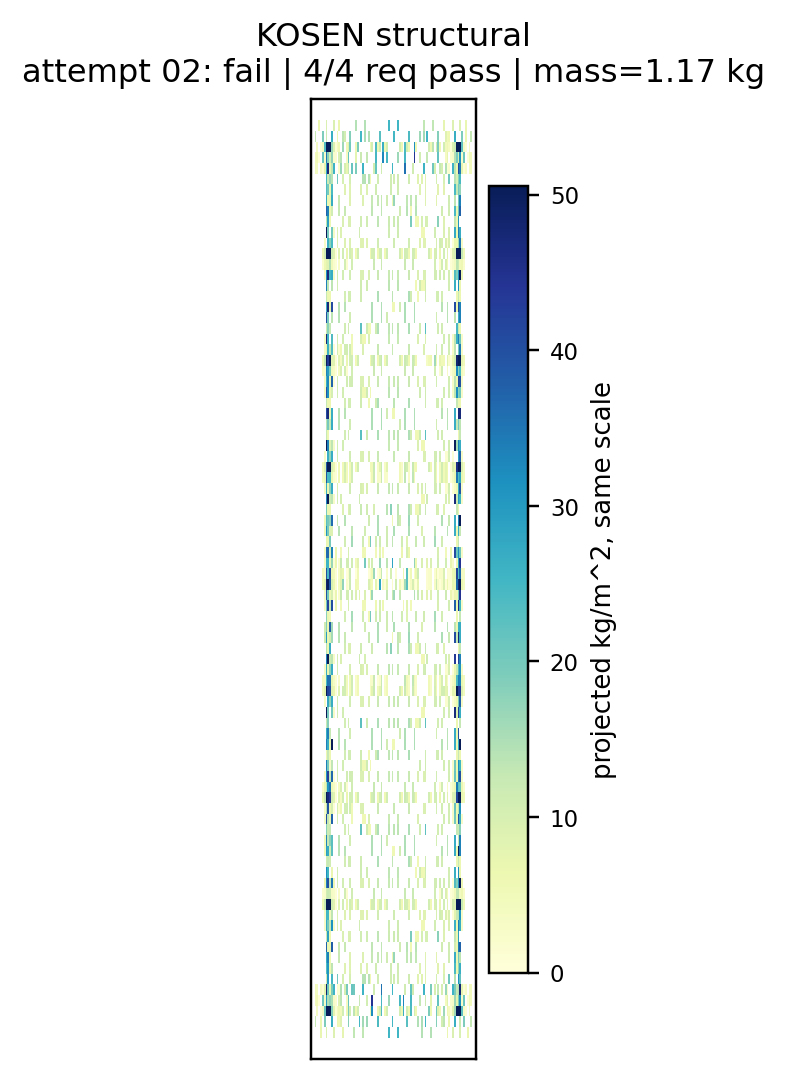} &
\includegraphics[width=0.175\linewidth]{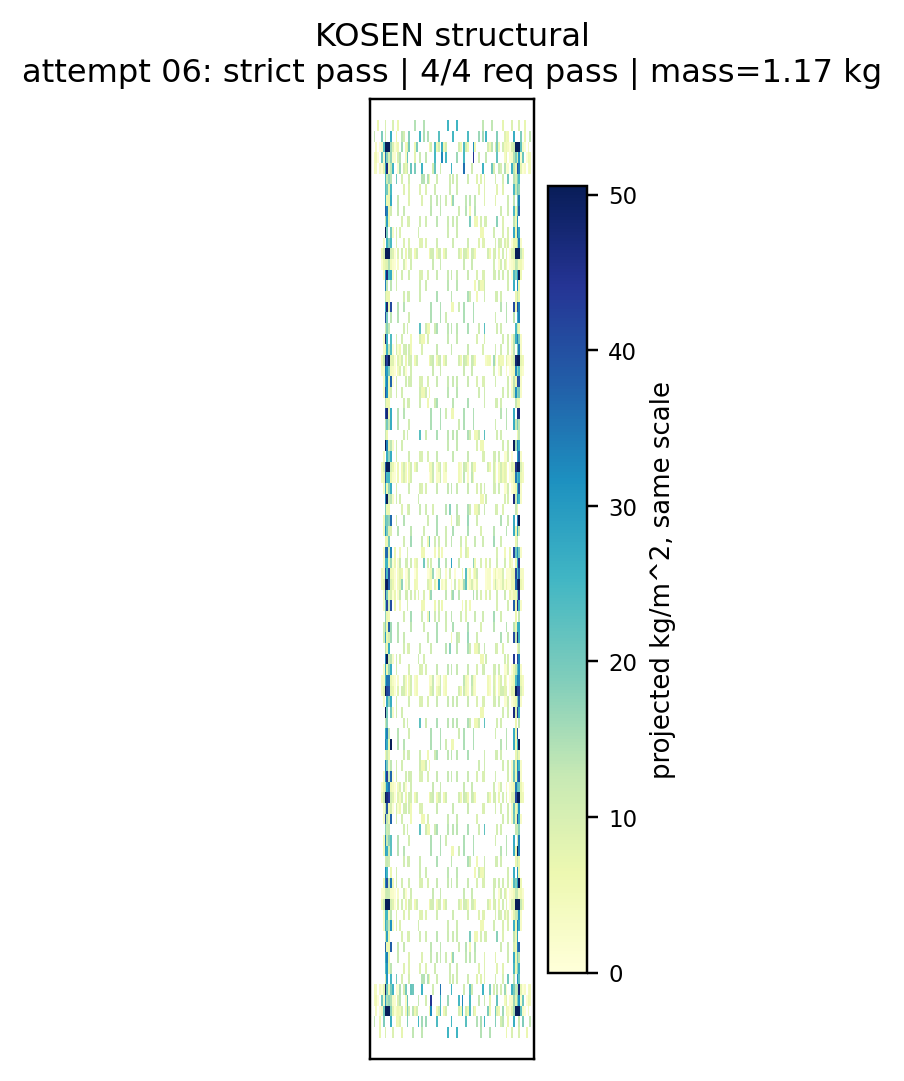} \\
\bottomrule
\end{tabular}
\caption{\footnotesize Strict-pass retries dominated by checker-contract repair. These artifacts already satisfy much of the underlying physics, but fail strict grading until the generated metadata exposes the required metric names, mass fields, selector bindings, or mesh-derived mass aliases.}
\label{fig:strict-pass-samples-contract}
\end{figure}

Figure~\ref{fig:compute-scaling} summarizes the nine artifacts that become strict passes in the longest \texttt{GPT-5.5/high} feedback run. Figures~\ref{fig:strict-pass-samples-structural}--\ref{fig:strict-pass-samples-contract} show the corresponding before/after images for all nine cases. Each row contains the earliest retained failing artifact and the selected passing retry: surface render before, surface render after, checker-derived field before, and checker-derived field after. Stress-driven cases use von Mises proxy fields; mass-driven or contract-driven cases use projected mass or density fields. The appendix figures are meant to make the repair mechanisms in Figure~\ref{fig:compute-scaling} concrete: some fixes are visible shape changes, but several pass flips are only visible in field maps or requirement bindings.

\paragraph{Structural retuning.}
The AISC 360-22 steel column is the clearest load-path change and corresponds to Figure~\ref{fig:compute-scaling}C. The failed retry is a slender weak-axis member that misses compression, bending, and service-deflection requirements; the passing retry becomes a four-chord braced box column, raising compression and bending capacity while reducing deflection below the limit without exceeding the weight cap. The HPVC roll-protection system starts from a structure that already satisfies stress and deflection checks but is far too heavy; the passing retry replaces the over-heavy solid surrogate with thinner rectangular members, preserving stiffness while dropping frame mass below the limit. The Teknofest agricultural UGV tool arm combines infrastructure repair, selector cleanup, and structural retuning: the passing retry rebuilds the arm as a coarse hollow box beam with broad root and tip selector faces, reducing stress, tip deflection, fatigue stress, and dry mass while meeting the modal-frequency requirement.

\paragraph{Simplification and hidden physical-property repair.}
The RoboMaster 17~mm launcher corresponds to Figure~\ref{fig:compute-scaling}B. The failed design contains detailed launcher geometry with stress, radial-growth, and meshing failures; the passing retry uses a simpler monolithic surrogate that meshes robustly, lowers peak stress, and keeps barrel displacement within tolerance. The FIA Article 253 rollcage follows the same broad pattern: the failed cage is dense, over-stressed, and numerically unstable, while the passing retry uses a lighter FIA-compatible tube layout and adds the tube and mount-compliance metadata needed by the checker. The ECSS spacecraft panel is different and corresponds to Figure~\ref{fig:compute-scaling}D. Its failed version already has large stress, buckling, and modal margins, but its areal density is more than an order of magnitude too high; the passing retry preserves the panel planform while switching to a lightweight sandwich-equivalent representation, reducing areal density to the allowable range.

\paragraph{Checker-contract repair.}
The ISO 10328 P5 prosthetic pylon is mostly a checker-contract repair with some visible mass trimming. The geometry changes from an adapter-heavy pylon to a trimmed-adapter, stronger-tube version, but the decisive fix is exposing the expected toe-force, heel-force, and fatigue-life metric aliases so the already-low-stress design can satisfy the ISO P5 checks under the mass limit. The AIJ structural design and KOSEN Dezacon structural design are even more contract-dominated. In both cases, displacement and buckling already pass by large margins, but strict grading fails because the checker cannot bind the mass requirement. The passing retries add mesh-derived mass, self-weight, and related aliases, exposing the already-acceptable mass to the requirement checker. These cases are why rendered-view feedback alone is insufficient: the failure is not the shape, but the bridge between the generated artifact and the evaluator's typed engineering contract.

\clearpage

\section{Rich-view render set}
\label{app:richview}

\begin{table}[t]
\centering
\caption{The ParaView 21-view render set used by the image feedback judge. All views share a fixed camera distance scaled by the bounding sphere of the part; close-ups and x-ray views additionally vary zoom and per body alpha.}
\label{tab:richview}
\small
\begin{tabular}{p{0.20\linewidth} c p{0.62\linewidth}}
\toprule
Group & Count & Views \\
\midrule
Axis-aligned and isometric & 12 & front, rear, left, right, top, bottom, iso, iso\_front\_right, iso\_rear\_right, iso\_rear\_left, iso\_front\_left, iso\_top\_front\_right \\
Close-ups (zoom $\times 1.45$ to $\times 1.8$) & 6 & front\_close, rear\_close, left\_close, right\_close, top\_close, iso\_close \\
X-ray (per body alpha blending) & 3 & iso\_xray, front\_xray, right\_xray \\
\bottomrule
\end{tabular}
\end{table}

\clearpage

\section{Benchmark detail: sources, distribution, schema}
\label{app:benchmarkdetail}

This section covers the authoring methodology and scrub pass (Section~\ref{app:bench:authoring}), the schema each brief follows (Section~\ref{app:bench:schema}), the source catalogs and per-catalog distribution (Section~\ref{app:bench:sources}), the distribution of pass/fail requirements across the pool (Section~\ref{app:bench:reqdist}), the criteria for selecting the curated 50-case benchmark from the 466-case pool (Section~\ref{app:bench:selection}), and three sample briefs (Section~\ref{app:bench:sample}).

\subsection{Authoring methodology and scrub pass}
\label{app:bench:authoring}

Each brief was authored through the same five-step workflow:
\begin{enumerate}
\item \textbf{Catalog identification.} A new catalog (e.g., a standards body, a supplier datasheet family, or a regional industrial archive) is added to the candidate pool, and each prospective case is assigned a difficulty tier (A+, A, B, C) reflecting how completely the underlying source already specifies the FEA pass/fail criteria.
\item \textbf{Brief drafting.} The narrative engineering brief is written from the source material, with all numeric limits (load magnitudes, material strengths, dimensional tolerances, safety factors) resolved inline at authoring time so the brief carries no dangling references to external documents.
\item \textbf{Requirement expansion.} The brief's pass/fail criteria are decomposed into an explicitly typed list of requirements (R1\dots Rn), each one tied to a load case and a numeric threshold.
\item \textbf{Checker validation.} Every candidate brief is run through the CalculiX harness against a hand-built reference geometry. The brief is only kept if every requirement parses, evaluates, and binds without skipping.
\item \textbf{Scrub pass before release.} A final pass strips identifying metadata and external dependencies; details follow.
\end{enumerate}

The scrub enforced two invariants. (i)~\emph{No source identity}: we removed every metadata field that could attribute a brief to a specific organization, competition, or human author, and rewrote the prompt text of 49 files whose narratives named specific competitions (e.g., Formula SAE, GrabCAD, Teknofest, Robocon, Hyperloop Pod, Solar Decathlon, DARPA, AFRL), replacing each with generic phrasing such as ``collegiate formula-class racecar benchmark.'' (ii)~\emph{No external dependency}: we removed 24 governing-standard URL fields and verified that every pass/fail criterion now carries its numeric threshold inline, except for criteria that are exempt by construction (material certificates, geometric checks, dimensional compliance). Internal filename and identifier fields were kept unchanged as opaque handles so internal traceability survives the public release without leaking source identity. The exact metadata and requirement fields touched by the scrub are listed in the file schema (Section~\ref{app:bench:schema}).

\subsection{File schema}
\label{app:bench:schema}

Each brief is a self-contained record pairing a prompt with a structured FEM-requirement set. The schema is fixed; the scrub pass in Section~\ref{app:bench:authoring} acts on a defined subset of these fields. The top-level fields are:
\begin{itemize}
  \item \texttt{full\_prompt}: the narrative engineering brief covering geometry, materials with full property tables, load cases (LC1\dots LCn with magnitudes), pass/fail criteria with derivations, solver expectations, and deliverables.
  \item \texttt{prompt.*}: a structured mirror of the prompt with explicit \texttt{geometric\_constraints}, \texttt{material(s)}, \texttt{load\_cases}, and \texttt{output\_format}.
  \item \texttt{requirements.pass\_fail\_criteria}: an ordered list of R1\dots Rn, each carrying \texttt{id}, \texttt{type} (one of \texttt{structural}, \texttt{vibration}, \texttt{thermal}, \texttt{fluid}, \texttt{radiation}, \texttt{buckling}, \texttt{dimensional}, \texttt{material\_compliance}, \texttt{geometric\_check}), \texttt{metric}, a numeric \texttt{limit\_*} field, \texttt{derivation}, \texttt{applies\_to}, and \texttt{operator}.
  \item \texttt{verification}: the primary FEA class, secondary classes, explicitly excluded classes, and \texttt{requires\_non\_fea\_solver} flags for analyses outside FEA scope (e.g., TID dose, trajectory simulation, topology optimization, LEFM, Paris-law fatigue).
  \item \texttt{notes.exclusions\_explained}: justifies each excluded analysis class so the harness does not silently mark missing physics as a failure.
  \item \texttt{eval\_coverage}: what the case is intended to test, used for stratified sampling.
  \item \texttt{source.*}: provenance metadata. \emph{Stripped at release time by the scrub pass} (Section~\ref{app:bench:authoring}). Only the catalog code (e.g., \texttt{pt}, \texttt{s}, \texttt{i\_eu}) and tier (A+/A/B/C) survive.
\end{itemize}
Numeric limits are written inline; standards (e.g., NASA-STD-5020, GEVS-STD-7000A) are cited by name only and the numeric values they imply are written into the brief.

\subsection{Source catalogs and per-catalog distribution}
\label{app:bench:sources}

The 466-case candidate pool (318 single-part briefs and 148 multi-part briefs) is drawn from sixteen catalogs, with source identity (author, host, URL) not persisted in the released specs.

\paragraph{Tier definitions.} Each case is tagged with a tier (A+/A/B/C) reflecting the strictness and self-containment of its underlying source.
\begin{itemize}
  \item \textbf{A+}: the source already publishes every numeric value the brief needs and the FEA pass/fail can be derived without authoring judgment (e.g., \href{https://standards.nasa.gov/standard/nasa/nasa-std-5020}{NASA-STD-5020} bolted-joint allowables, with exact load factors and thresholds spelled out in the standard).
  \item \textbf{A}: source has clear specs but one or two parameters require modest interpretation to bind (most engineering standards and curated supplier datasheets fall here, since the source typically fixes material and geometry but leaves the specific load magnitudes to the engineer).
  \item \textbf{B}: source provides a domain-realistic problem statement in which several parameters require authoring judgment to make the brief evaluable (e.g., competition rulebooks that fix the envelope and a survival criterion but leave the load derivation to the team).
  \item \textbf{C}: source provides framing only, with the numerical content fully assembled by the authors against domain conventions (mostly the open-ended intercollegiate cases that ship as design briefs without a published numeric rubric).
\end{itemize}
Tier influences both inclusion in the curated 50-case benchmark (Section~\ref{app:bench:selection}, where A+/A cases are over-sampled because their pass/fail rubrics are most defensible) and the level of scrutiny in the scrub pass (Section~\ref{app:bench:authoring}, where C-tier cases were hand-checked against domain conventions before release because there is no authoritative source to reconcile against).

\paragraph{Catalog inventory.} The catalogs span: patents and supplier datasheets (PT, e.g., GE jet engine bracket benchmark\footnote{\url{https://grabcad.com/challenges/ge-jet-engine-bracket-challenge}}, Hilti\footnote{\url{https://www.hilti.com}} Profis HIT-HY200, SKF\footnote{\url{https://www.skf.com}} and Schaeffler\footnote{\url{https://www.schaeffler.com}} bearing housings, Bosch Rexroth\footnote{\url{https://www.boschrexroth.com}} linear guide, PennEngineering\footnote{\url{https://www.pemnet.com}} PEM insert, Heli-Coil\footnote{\url{https://www.stanleyengineeredfastening.com/brands/heli-coil}} technical manual, 3M VHB\footnote{\url{https://www.3m.com}} structural glazing, Henkel Loctite\footnote{\url{https://www.henkel-adhesives.com}} structural adhesive, W\"urth\footnote{\url{https://www.wuerth.com}} fastener handbook, MISUMI\footnote{\url{https://www.misumi-ec.com}} aluminum frame bracket); engineering standards (S, e.g., NASA-STD\footnote{\url{https://standards.nasa.gov}}, ECSS\footnote{\url{https://ecss.nl}}, AISC\footnote{\url{https://www.aisc.org}}, MIL-STD\footnote{\url{https://quicksearch.dla.mil}}, FIA Art.253\footnote{\url{https://www.fia.com/sport/regulations}}, Cal Poly CDS\footnote{\url{https://www.cubesat.org}}); regional industrial catalogs by country (JP Japan, KR Korea, CN China, SA aerospace, OC Oceania, AF Africa, NG humanitarian / needs-grade, MD medical, SP showcase, E energy, HV heavy vehicles, DF defense); the A-series and the I-series, both detailed below.

\paragraph{A-series (A1--A45).} A foundational catalog of 45 reference cases drawn from open CAD datasets and well-known engineering benchmark briefs, with deliberate emphasis on aerospace and student-launch hardware. Examples include EuRoC\footnote{\url{https://www.euroc.pt}} pressure vessels, IREC\footnote{Intercollegiate Rocket Engineering Competition: \url{https://www.soundingrocket.org}} composite-overwrap pressure vessels (COPVs), NASA Student Launch\footnote{\url{https://www.nasa.gov/learning-resources/nasa-student-launch/}} airframes, Cal Poly 1U CubeSat chassis, AIAA Design-Build-Fly\footnote{\url{https://www.aiaa.org/dbf}} wings, SAE Aero Design\footnote{\url{https://www.sae.org/attend/student-events/sae-aero-design}} landing gear, IMechE UAS Challenge\footnote{\url{https://www.imeche.org/events/challenges/uas-challenge}} airframes, and AISC SSBC\footnote{\url{https://www.aisc.org/education/university-programs/student-steel-bridge-competition/}} structural-steel bridge variants. The A-series functions as the ``known'' anchor of the pool: every A-series brief has a publicly available reference design or rubric, which makes it useful both for sanity-checking the harness during authoring and for cross-referencing against prior literature.

\paragraph{I-series (intercollegiate).} A regional catalog of intercollegiate engineering-competition briefs split by world region, with sub-codes covering East Asia (\texttt{i\_ea}), Europe (\texttt{i\_eu}), Latin America (\texttt{i\_la}), Middle East (\texttt{i\_me}), South Asia (\texttt{i\_sa}), Oceania (\texttt{i\_oc}), and Africa (\texttt{i\_af}). Representative competitions include ABU Robocon\footnote{\url{https://aburobocon2025.com}} and KOSEN Robocon\footnote{\url{https://official-robocon.com/kosen/}} (East Asia); RoboMaster\footnote{\url{https://www.robomaster.com/en-US}} robotics competitions (East Asia); FSAE Japan\footnote{\url{https://www.jsae.or.jp/formula/en/}} and Formula Student China\footnote{\url{https://www.formulastudent.cn}} (East Asia); ESA REXUS / BEXUS sounding-rocket and balloon programs\footnote{\url{https://rexusbexus.net}} and ESA CanSat Europe\footnote{\url{https://www.esa.int/Education/CanSat}} (Europe); Formula Student Germany\footnote{\url{https://www.formulastudent.de}} (Europe); Baja SAE Brasil\footnote{\url{https://saebrasil.org.br}} (Latin America); Teknofest\footnote{\url{https://www.teknofest.org/en/}} subsystems including airframes, AUV pressure hulls, and tarim (agricultural) UGVs (Middle East); Formula Bharat\footnote{\url{https://www.formulabharat.com}}, SUPRA SAEINDIA\footnote{\url{https://www.saeisst.in/supra}}, INSPACE CanSat India\footnote{\url{https://www.inspaceindia.in}}, and Aerothon\footnote{\url{https://www.aerothon.in}} (South Asia). I-series cases tend to be open-ended design briefs without published numeric rubrics (most are tier-C in our tier scheme), so they primarily test the model's ability to generate engineering-credible CAD against domain-conventional constraints and not to match a single curated reference.

\begin{figure}[h]
\centering
\includegraphics[width=\linewidth]{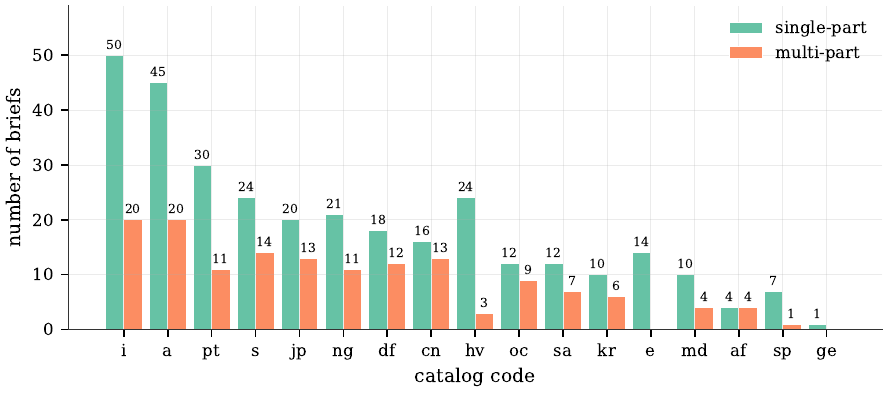}
\caption{Candidate-pool brief distribution by catalog, single-part vs multi-part. The intercollegiate catalog (\texttt{i}) and the foundational A-series (\texttt{a}) account for the largest share of the pool; engineering standards (\texttt{s}) and patents/datasheets (\texttt{pt}) provide the bulk of the strict-spec briefs.}
\label{fig:catdist}
\end{figure}

Per-catalog counts are given in Table~\ref{tab:catalogdist} and visualized in Figure~\ref{fig:catdist}.

\begin{figure}[h]
\centering
\includegraphics[width=\linewidth]{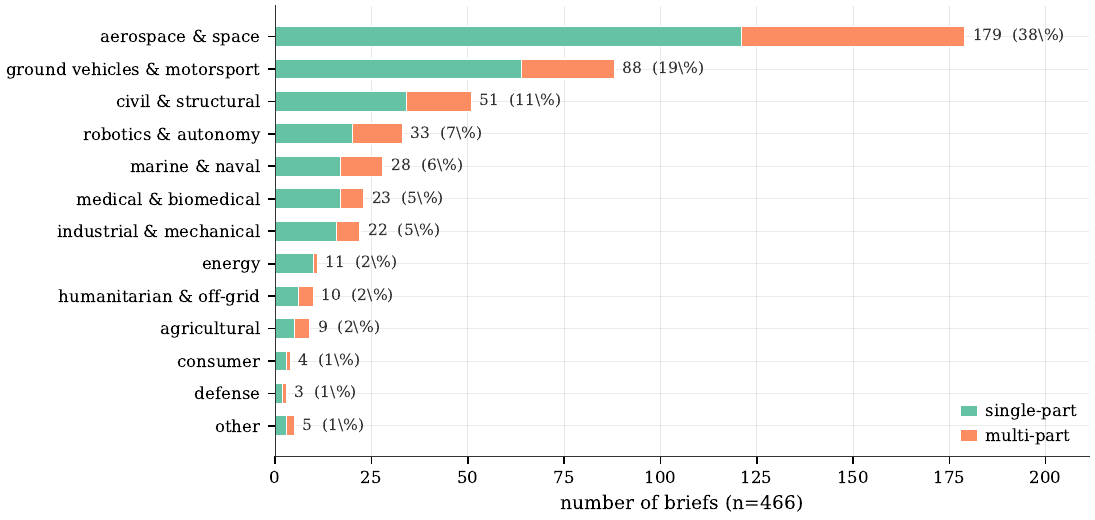}
\caption{Per-item engineering-domain distribution across the 466-case candidate pool, with the raw \texttt{domain} field grouped into thirteen broad buckets. Each bar is split into single-part and multi-part segments; the right-of-bar number is the total count and percentage of the pool. Aerospace and ground-vehicle cases together account for over half the pool, but every bucket is represented in both subsets, which is what lets the curated 50-case benchmark stay diverse despite the catalog skew toward intercollegiate and A-series sources.}
\label{fig:domaindist}
\end{figure}

\paragraph{Distribution by physical domain.} Beyond catalog provenance, every brief carries a \texttt{domain} field that classifies the part by engineering domain (e.g., aerospace structural, civil structural, mechanical actuator, biomedical implant, thermal-management housing). Figure~\ref{fig:domaindist} shows the per-item domain distribution across the pool.

\begin{table}[h]
\centering
\caption{Per-catalog distribution of candidate briefs in the pool from which \newbench{} is drawn.}
\label{tab:catalogdist}
\small
\begin{tabular}{l l r r r}
\toprule
Code & Domain & Single & Multi & Total \\
\midrule
\texttt{i}  & intercollegiate competitions       & 50  & 20  & 70  \\
\texttt{a}  & A-series (A1--A45) original         & 45  & 20  & 65  \\
\texttt{pt} & patents and supplier datasheets    & 30  & 11  & 41  \\
\texttt{s}  & engineering standards              & 24  & 14  & 38  \\
\texttt{hv} & heavy vehicles                     & 24  & 3   & 27  \\
\texttt{ng} & humanitarian / needs-grade         & 21  & 11  & 32  \\
\texttt{jp} & Japan industrial                   & 20  & 13  & 33  \\
\texttt{df} & defense                            & 18  & 12  & 30  \\
\texttt{cn} & China industrial                   & 16  & 13  & 29  \\
\texttt{e}  & energy                             & 14  & 0   & 14  \\
\texttt{sa} & aerospace                          & 12  & 7   & 19  \\
\texttt{oc} & Oceania industrial                 & 12  & 9   & 21  \\
\texttt{md} & medical                            & 10  & 4   & 14  \\
\texttt{kr} & Korea industrial                   & 10  & 6   & 16  \\
\texttt{sp} & showcase                           & 7   & 1   & 8   \\
\texttt{af} & Africa industrial                  & 4   & 4   & 8   \\
\texttt{ge} & generic                            & 1   & 0   & 1   \\
\midrule
\textbf{Total} &                                & \textbf{318} & \textbf{148} & \textbf{466} \\
\bottomrule
\end{tabular}
\end{table}

\subsection{Distribution by requirement type and per-brief count}
\label{app:bench:reqdist}

\begin{figure}[h]
\centering
\includegraphics[width=0.85\linewidth]{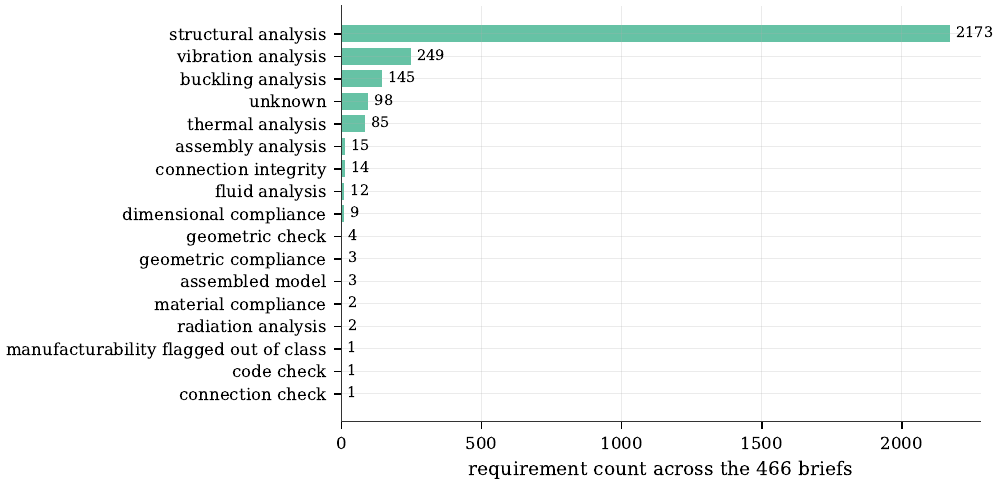}
\caption{Distribution of requirement \texttt{type} across all pass/fail criteria in the 466-case pool. Structural-analysis criteria dominate; buckling, vibration, thermal, dimensional, geometric, and material-compliance checks each contribute a meaningful share. The two smallest types (\texttt{fluid}, \texttt{radiation}) sit outside CalculiX's scope and are tracked as future-work analyses.}
\label{fig:reqdist}
\end{figure}

Across the 466 briefs, every pass/fail criterion is tagged with a discrete \texttt{type}. Figure~\ref{fig:reqdist} breaks down requirement type frequency across the whole pool: structural analysis dominates, with substantial coverage of buckling, vibration, thermal, dimensional, geometric, and material-compliance checks. The pool carries 2{,}817 typed requirements in total, with single-part briefs averaging 5.7 requirements each (median 6, range 3--10) and multi-part briefs averaging 6.8 (median 7, range 3--10); multi-part briefs skew slightly higher because every assembly interface adds at least one weld or mate check. \emph{A subset of types is not yet evaluable by the CCX harness}: \texttt{fluid} requirements (CFD, e.g., aerodynamic load resolution from a wing's flow field) and \texttt{radiation} requirements (e.g., total ionizing dose, neutron flux) sit outside CalculiX's scope and are flagged in each brief via the \texttt{requires\_non\_fea\_solver} field; we keep them in the schema for completeness and tracking, and treat their integration with an external CFD/transport solver as future work.

\subsection{Selection of the 50-case curated benchmark}
\label{app:bench:selection}

\begin{figure}[h]
\centering
\includegraphics[width=\linewidth]{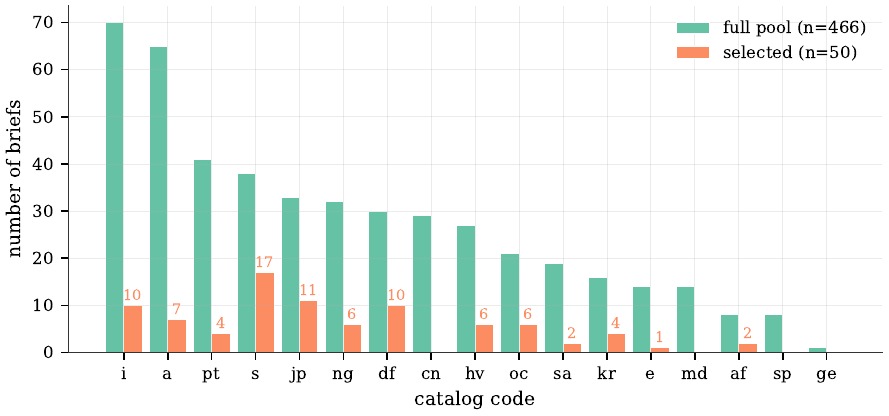}
\caption{Catalog coverage of the curated 50-case benchmark against the full 466-case candidate pool. Bars are the pool count; red overlay is the count selected for \newbench{}. The selection over-samples engineering-standards (\texttt{s}) and aerospace (\texttt{sa}) catalogs because those briefs exercise the strictest pass/fail rubrics, and samples the I-series and A-series lightly relative to their pool share to keep the curated benchmark from being dominated by intercollegiate cases.}
\label{fig:selvspool}
\end{figure}

\newbench{}'s 50 cases (20 single-part, 30 multi-part) are sampled from the 466-case pool by stratifying on three axes. (i)~\emph{Domain coverage}: we required at least one case from every catalog where the candidate pool exceeded five briefs, prioritizing the high-tier (A+/A) cases within each catalog. (ii)~\emph{Analysis-type coverage}: we required the curated set to exercise every \texttt{*STATIC}, \texttt{*FREQUENCY}, \texttt{*BUCKLE}, \texttt{*DYNAMIC}, and \texttt{*HEAT TRANSFER} card under the CalculiX backend, so the harness is end-to-end exercised by the 50 cases alone. (iii)~\emph{Difficulty spread}: within each catalog we kept a mix of monolithic single-part briefs (which mostly exercise R1--R5-type structural / buckling checks) and multi-part briefs that introduce assembly interface checks (weld DCR, mating clearance, deformed envelope). Figure~\ref{fig:selvspool} compares the curated 50 against the full pool by catalog.

\subsection{Sample briefs}
\label{app:bench:sample}

Three briefs illustrate the schema across different physics regimes. Each shows an excerpted prompt and the tagged pass/fail requirements.

\begin{tcolorbox}[reqbox, title={Brief A: GE jet engine bracket (aerospace structural, single-part)}]
\textbf{Prompt (excerpt).} \itshape Design a single-part, monolithic titanium mounting bracket that connects a jet engine casing to an external accessory, transferring load without permanent deformation under transient post-FBO (Fan Blade Out) conditions. The objective is to minimize mass while passing four independent static load cases. The bracket is a single solid body with no fasteners, inserts, welded joints, or bonded joints. Envelope: 203~mm $\times$ 152~mm $\times$ 102~mm. Top face is a planar flange against the engine casing with four 12.7~mm bolt holes on a 76.2~$\times$~76.2~mm rectangular pattern; bottom face carries an accessory pin\dots\upshape

\medskip
\textbf{Pass/fail requirements.}
\begin{lstlisting}
- id: R1, type: structural_analysis,
  metric: max_von_mises_stress, op: "<=", limit_MPa: 633,
  derivation: ultimate_strength/1.5 (CS-25 airframe ultimate SF),
  applies_to: [LC1, LC2, LC3, LC4]
- id: R2, type: structural_analysis,
  metric: max_von_mises_stress, op: "<=", limit_MPa: 748,
  derivation: yield_strength/1.15,
  applies_to: [LC1, LC2, LC3, LC4]
  note: R1 binds since 633 < 748; R2 retained for traceability
- id: R3, type: structural_analysis,
  metric: max_displacement_at_pin_hole_center, op: "<=", limit_mm: 1.0,
  applies_to: [LC1, LC2, LC3, LC4]
- id: R4, type: structural_analysis,
  metric: mass, op: "<=", limit_kg: 1.0,
  secondary: at least 40% lighter than bulk-block baseline,
  applies_to: [design]
- id: R5, type: buckling_analysis,
  metric: first_mode_load_factor, op: ">=", limit: 2.0,
  applies_to: [LC3]
  rationale: LC3 has the largest compressive component
\end{lstlisting}
\end{tcolorbox}

\begin{tcolorbox}[reqbox, title={Brief B: NASA-HDBK-7005 random-vibration satellite enclosure (aerospace structural, multi-part)}]
\textbf{Prompt (excerpt).} \itshape Design a 6U small-satellite equipment box per NASA-HDBK-7005 (2001), delivered as a multi-part assembly for random vibration, sine, and shock qualification (GSFC-STD-7000 / GEVS reference). Assembly: a milled Al 6061-T6 enclosure (200~$\times$~150~$\times$~70~mm, nominal 2~mm wall, flange with 6~M4 holes on a 180~mm bolt circle)\dots\upshape

\medskip
\textbf{Pass/fail requirements.}
\begin{lstlisting}
- id: R1, type: vibration_analysis,
  metric: first_natural_frequency_flange_constrained, op: ">=", limit_Hz: 140,
  applies_to: [constrained_modal]
- id: R2, type: vibration_analysis,
  metric: random_3sigma_wall_stress, op: "<=", limit_MPa: 221.4,
  applies_to: [LC1]
- id: R3, type: vibration_analysis,
  metric: random_3sigma_fastener_ligament_stress, op: "<=", limit_MPa: 220.8,
  applies_to: [LC1]
- id: R4, type: vibration_analysis,
  metric: sine_peak_stress, op: "<=", limit_MPa: 220.8,
  applies_to: [LC2]
- id: R5, type: vibration_analysis,
  metric: shock_SRS_peak_stress, op: "<=", limit_MPa: 221.4,
  applies_to: [LC3]
- id: R6, type: structural_analysis,
  metric: qs_30g_max_stress, op: "<=", limit_MPa: 220.8,
  applies_to: [LC4]
- id: R7, type: structural_analysis,
  metric: empty_enclosure_mass, op: "<=", limit_g: 800,
  applies_to: [assembly]
\end{lstlisting}
\end{tcolorbox}

\begin{tcolorbox}[reqbox, title={Brief C: Hilti HIT-HY200 anchored steel baseplate (civil structural, multi-part)}]
\textbf{Prompt (excerpt).} \itshape Design a bolted/adhesive-anchored steel baseplate assembly anchoring a tubular steel column into a cracked C20/25 normal-weight concrete slab, delivered as an explicit multi-body system. The plate is S355 steel 300~$\times$~300~mm ($\geq$~20~mm thick) with four 18~mm anchor clearance holes on a 200~$\times$~200~mm pattern\dots\upshape

\medskip
\textbf{Pass/fail requirements.}
\begin{lstlisting}
- id: R1, type: structural_analysis,
  metric: per_anchor_tension_demand, op: "<=", limit_kN: 37,
  applies_to: [LC1]
- id: R2, type: structural_analysis,
  metric: per_anchor_shear_demand, op: "<=", limit_kN: 43,
  applies_to: [LC1]
- id: R3, type: structural_analysis,
  metric: combined_tension_shear_interaction, op: "<=", limit: 1.0,
  applies_to: [LC1]
- id: R4, type: structural_analysis,
  metric: concrete_cone_utilization, op: "<=", limit: 1.0,
  applies_to: [LC1]
- id: R5, type: structural_analysis,
  metric: plate_max_bending_stress, op: "<=", limit_MPa: 319.5,
  applies_to: [LC1]
- id: R6, type: structural_analysis,
  metric: plate_deflection_across_column_footprint, op: "<=", limit_mm: 1.0,
  applies_to: [LC1]
- id: R_asm1, type: connection_integrity,
  metric: connection_DCR, op: "<=", limit: 1.0,
  applies_to: [LC1]
- id: R_asm2, type: connection_integrity,
  metric: fillet_weld_DCR, op: "<=", limit: 1.0,
  applies_to: [LC1]
\end{lstlisting}
\end{tcolorbox}

\clearpage

\section{Release, Reproducibility, Compute, and Impact}
\label{app:release-repro}

\paragraph{Release.}
We release \newbench{} as an anonymized supplemental zip at submission time. The release contains the authored benchmark briefs, requirement metadata, evaluation harness, and scripts needed to run the reported checks. The new benchmark assets and code authored for this paper are released under the MIT License; third-party datasets, tools, and services retain their original licenses and terms, summarized in Table~\ref{tab:asset-licenses}. We will add an anonymized GitHub mirror when available during review, and a public GitHub repository after deanonymization.

\paragraph{Reproduction environment and commands.}
The local pipeline targets Python~3.12 and the package set in \texttt{requirements.txt}, including \texttt{cadquery==2.7.0}, \texttt{openai==2.29.0}, \texttt{litellm==1.82.4}, \texttt{langgraph==1.1.3}, and \texttt{google-genai==1.68.0}. The CCX grader is \texttt{scripts/ccx\_eval/grade\_ccx.py}. By default it uses \texttt{CCX=/opt/homebrew/bin/ccx\_2.22}, \texttt{EVAL\_PYTHON=/opt/anaconda3/envs/cadquery/bin/python}, and the Gmsh Python API if \texttt{GMSH} is unset. Gmsh uses HXT tetrahedral meshing with curvature sizing and mesh-size bounds of 1.0--50.0~mm; rich-view renders use ParaView 6.1.0.

\begin{lstlisting}[basicstyle=\ttfamily\scriptsize,breaklines=true]
pip install -r requirements.txt
python scripts/fire_ccx_supervised_loop.py \
  --backend codex --set all --scope curated \
  --model gpt-5.5 --reasoning high \
  --max-attempts 15 --jobs 8 \
  --timeout-model 2400 --timeout-eval 900 \
  --skill-mode cad --feedback-mode deep-feedback \
  --require-rich-view \
  --out-root runs/<run_name>
python scripts/eval_codex_ccx_runs.py \
  --set single --scope curated --run-dir <run_dir> \
  --out-root <eval_dir> --jobs 8 --timeout 900
CCX=/opt/homebrew/bin/ccx_2.22 \
EVAL_PYTHON=/opt/anaconda3/envs/cadquery/bin/python \
python scripts/ccx_eval/grade_ccx.py <case_workdir>
\end{lstlisting}

\paragraph{Compute and statistical uncertainty.}
All local orchestration, CAD execution, meshing, rendering, and FEA jobs reported in this paper were run on CPU workers; no local GPU workers were used. The main long \texttt{GPT-5.5/high} feedback run used \texttt{--jobs 8}, a 2400-second model timeout, a 900-second evaluation timeout, and ran for 359{,}525.79 wall-clock seconds from 2026-05-02 to 2026-05-06. The paper reports aggregate metrics on fixed benchmark subsets rather than repeated-run estimates. We do not report confidence intervals, error bars, or significance tests because repeating every full agent run was cost-prohibitive; this should be read as an uncertainty limitation on the empirical comparisons.

\paragraph{Broader impacts and responsible use.}
The main positive impact is a more engineering-grounded way to evaluate and improve CAD agents: generated artifacts are checked against explicit physical and geometric requirements rather than only appearance or reference similarity. This can help researchers study failures before such systems are used in design workflows. The main risks are over-trust in generated CAD, propagation of mechanically invalid load paths, and dual-use mechanical design automation. \newbench{} and the released harness are evaluation assets, not a certified design tool. Generated artifacts should not be used for safety-critical, regulated, or manufactured designs without independent professional engineering review, solver validation, and domain-specific certification. The benchmark scrub pass in Appendix~\ref{app:bench:authoring} removes source identities and external dependencies, and the release focuses on prompts, metadata, and evaluators rather than a deployable autonomous design service.

\begin{table}[h]
\centering
\caption{Existing assets, tools, and service terms used by the paper.}
\label{tab:asset-licenses}
\scriptsize
\begin{tabular}{p{0.22\linewidth} p{0.26\linewidth} p{0.43\linewidth}}
\toprule
Asset & License or terms & Source / note \\
\midrule
\texttt{S2O} & Code: MIT. Data: Hugging Face ``other'' license that inherits PartNet-Mobility, HSSD (CC BY-NC 4.0), ABO, and 3D-FUTURE terms. & \href{https://github.com/3dlg-hcvc/s2o}{code repository}; \href{https://huggingface.co/datasets/3dlg-hcvc/s2o}{dataset card}. \\
\texttt{Fusion\,360 Gallery Assembly} & Autodesk dataset license; non-commercial research use; no redistribution of the full dataset. & \href{https://github.com/AutodeskAILab/Fusion360GalleryDataset/blob/master/LICENSE.md}{dataset license}. \\
CadQuery & Apache License 2.0. & \href{https://github.com/CadQuery/cadquery}{project repository}. \\
CalculiX & GNU GPL version 2 or later. & \href{https://www.calculix.de}{project website}. \\
Gmsh & GNU GPL version 2 or later, with the Gmsh linking exception. & \href{https://gmsh.info}{project website}. \\
ParaView & BSD 3-Clause. & \href{https://www.paraview.org/license/}{license page}. \\
OpenAI Codex & OpenAI Service Terms; code-generation output may be subject to third-party licenses. & \href{https://openai.com/policies/service-terms/}{service terms}. \\
Claude Code & Anthropic Commercial or Consumer Terms, plus Anthropic Usage Policy. & \href{https://code.claude.com/docs/en/legal-and-compliance}{Claude Code legal page}. \\
\bottomrule
\end{tabular}
\end{table}

\clearpage

\section{Full blueprint for Brief~\ref{brief:bajaframe}}
\label{app:fullblueprint}

The following schema-v4 blueprint corresponds to the Baja FSAE tubular space frame brief introduced in Section~\ref{sec:problem:gap}. The full frame decomposes into eight named tube families (\texttt{main\_hoop}, \texttt{front\_hoop}, \texttt{lower\_frame}, \texttt{side\_impact}, \texttt{roof\_bracing}, \texttt{front\_nose}, \texttt{rear\_bay}, \texttt{lateral\_bracing}); for compactness we show three representative parts.

\begin{lstlisting}
assembly_schema_version: 4
metadata:
  brief_id: brief:bajaframe
  units: mm
  coordinate_system: { x: forward, y: driver_right, z: up }
  material:
    name: AISI 1018 DOM
    yield_strength_MPa: 370
    safety_factor: 1.5
parts:
  - name: main_hoop
    geometry_definition:
      bounding_envelope:
        x: [-460, 460]
        y: [-50, 50]
        z: [0, 1100]
      support_zones:
        - name: weld_pad_left
          plane: { normal: "+y", offset: 50.0 }
          footprint: { x_span: [-450, -430], z_span: [1080, 1100] }
        - name: weld_pad_right
          plane: { normal: "-y", offset: -50.0 }
          footprint: { x_span: [430, 450], z_span: [1080, 1100] }
    construction_units:
      - id: hoop_tube
        role: additive
        envelope:
          must_fit_inside: { x: [-460, 460], y: [-50, 50], z: [0, 1100] }
        constructive_primitives:
          - op: cylinder
            axis: y
            radius_outer: 12.7
            wall_thickness: 3.05
            sweep_path: main_hoop_centerline
      - id: corner_fillet
        role: modifier
        constructive_primitives:
          - op: fillet_hint
            edge_selector: hoop_tube.outer_edges_top
            radius: 4.0
    acceptance_claims:
      - { id: R1, metric: tube_OD_mm, operator: "==", value: 25.4 }
      - { id: R4, metric: first_mode_load_factor_LC4, operator: ">=", value: 1.5 }

  - name: side_impact
    geometry_definition:
      bounding_envelope:
        x: [-1000, 1000]
        y: [-460, 460]
        z: [400, 600]
    construction_units:
      - id: sim_tube_left
        role: additive
        envelope:
          must_fit_inside: { x: [-1000, 1000], y: [-460, -430], z: [400, 600] }
        constructive_primitives:
          - op: cylinder
            axis: x
            radius_outer: 12.7
            wall_thickness: 3.05
            sweep_path: sim_centerline_left
      - id: sim_tube_right
        role: additive
        envelope:
          must_fit_inside: { x: [-1000, 1000], y: [430, 460], z: [400, 600] }
        constructive_primitives:
          - op: cylinder
            axis: x
            radius_outer: 12.7
            wall_thickness: 3.05
            sweep_path: sim_centerline_right
    acceptance_claims:
      - { id: R3, metric: helmet_location_deflection_mm, operator: "<=", value: 25 }

  - name: suspension_pickup_tab_fr
    geometry_definition:
      bounding_envelope:
        x: [1180, 1280]
        y: [-170, 170]
        z: [220, 630]
    construction_units:
      - id: tab_plate
        role: additive
        envelope:
          must_fit_inside: { x: [1180, 1280], y: [-170, -150], z: [220, 280] }
        constructive_primitives:
          - op: extrude_polygon
            polygon_2d: [[0, 0], [40, 0], [40, 30], [0, 30]]
            extrude_axis: y
            extrude_length: 6.0
      - id: bolt_hole
        role: modifier
        constructive_primitives:
          - op: subtract_cylinder
            axis: y
            radius: 5.0
            center_2d: [20, 15]
    acceptance_claims:
      - { id: R_asm1, metric: weld_joint_DCR_max, operator: "<=", value: 1.0 }
\end{lstlisting}

\end{document}